\documentclass[aps,prb,reprint,showpacs,showkeys,superscriptaddress,nofootinbib,floatfix]{revtex4-1}
\usepackage[english]{babel}
\usepackage{hyperref}
\usepackage{amsmath}
\usepackage{amsfonts}
\usepackage{amssymb}
\usepackage{graphicx}
\usepackage{subfigure}
\usepackage{feynmp}

\usepackage{color}

\def\bs{{\bf S}}
\def\bk{{\bf k}}

\def\bq{{\bf q}}
\def\bQ{{\bf Q}}
\def\b0{{\bf 0}}
\def\br{{\bf r}}

\def\dag{\dagger}

\def\bra{\langle}
\def\ket{\rangle}

\def\vev#1{\langle{#1}\rangle}
\def\emin{\epsilon_{\rm min}}
\def\non{\nonumber\\}

\begin{document}
\title{Magnon condensation with finite degeneracy  on the triangular lattice}
\date{\today}
\author{Giacomo Marmorini}
\email{giacomo@riken.jp}
\affiliation{Condensed Matter Theory Laboratory, RIKEN, 2-1 Hirosawa, Wako, Saitama 351-0198, Japan} 
\affiliation{Research and Education Center for Natural Sciences, Keio University, 4-1-1 Hiyoshi, Kanagawa 223-8521, Japan}
\author{Tsutomu Momoi}
\affiliation{Condensed Matter Theory Laboratory, RIKEN, 2-1 Hirosawa, Wako, Saitama 351-0198, Japan}
\affiliation{RIKEN Center for Emergent Matter Science, 2-1 Hirosawa, Wako, Saitama 351-0198, Japan}

\begin{abstract}
We study the spin 1/2 triangular-lattice $J_1$-$J_2$-$J_3$
antiferromagnet close to the saturation field using the dilute Bose gas theory,
where the magnetic
structure is determined by the condensation of magnons.
We focus on
the case of ferromagnetic $J_1$ and antiferromagnetic {$J_2,J_3$}, that is particularly rich because
frustration effects allow the single-magnon energy dispersion to have six-fold degenerate minima
at incommensurate momenta.
Our calculation also includes an interlayer coupling $J_0$, which covers both antiferromagnetic
and ferromagnetic cases including negligibly small regime (two-dimensional case).
%,
% that  makes the system three-dimensional, but
% can be taken as small as $10^{-4}$ (in units of $J_1$),
%in which case the system is effectively two-dimensional.
%
Besides the spiral and fan phases, we find a new double-$q$ phase
(superposition of two modes), dubbed ``${\bf Q}_0$-${\bf Q}_1$''  (or simply ``01'') phase,
that enjoys a new type of multiferroic character. Certain phase boundaries have a singular $J_0$ dependence
for $J_0\to 0$, implying that even a very small interlayer coupling drastically changes the ground state.
A mechanism for this singularity is presented.
 Moreover, in some regions of the parameter space, we show that a dilute gas of magnons
 can not be stable, and phase separation (corresponding to a magnetization jump) is expected.
 In the $J_1$-$J_2$ model  ($J_3=0$), formation of two-magnon bound states is observed, which
 can lead to a quadrupolar (spin-nematic) ordered phase.
Exact diagonalization analysis is also applied to the search of
bound states.
\end{abstract}

\pacs{75.10.Jm,75.30.Kz,75.85.+t}
%
%\keywords{magnon BEC}

\maketitle

\section{Introduction}
Frustrated spin systems are privileged hosts of exotic phases of matter.
Quantum spin liquids,
quantum spin nematics, and topological spin textures (such as skyrmion and vortex crystals) among others
have recently attracted a lot of theoretical and experimental interest.
In theoretical analysis, however, fully quantum mechanical treatments of these systems present
in general huge difficulties, most notoriously for the sign problem of quantum Monte-Carlo simulations.
It is therefore crucial to establish and develop fully quantum methods of exploring at least some parts of
the magnetic phase diagram.

Since the pioneering work of Batyev and Braginskii\cite{batyev1984antiferromagnet},
the dilute Bose gas theory of magnons near saturation field
%the Bose-Einstein condensation of magnons (magnon BEC)
has become one of the few approaches that deal with magnetic systems in a fully quantum-mechanical fashion.
In this theory, magnetic systems are mapped to interacting hard-core Bose gas of magnons\cite{1956PThPh..16..569M}
and the magnetic state near saturation is described as a dilute condensed Bose gas. %of magnons.
%, if only in a specific regime, namely near the saturation magnetic field.
With the work of Nikuni and Shiba\cite{nikuni1995hexagonal} on the prototypical triangular Heisenberg
antiferromagnet it was made clear that frustration can induce various types of Bose-Einstein condensation (BEC),
hence new magnetic phases, thanks to the more complex low-energy structure of magnons and their interaction.
These states are in general characterized by the coherent superposition of one or more spirals,
from which the terminology single-$q$ and multiple-$q$ states comes.

While the triangular-lattice Heisenberg antiferromagnets (and helimagnets in general\cite{ueda2009magnon})
near saturation can accommodate only single-$q$ (spiral) or coplanar double-$q$ (fan) phases,
here we are interested in new kinds of multiple-$q$ phases that appears due to condensation of
magnons at (unusual) multiple momenta.
In this respect, a necessary condition is a high degeneracy of inequivalent  single-magnon energy minima
(in momentum space), which is typically brought about  by competing exchange interactions.
For this purpose, we start in this paper from the triangular-lattice $J_1$-$J_2$-$J_3$ model
with ferromagnetic $J_1$
and antiferromagnetic $J_2,J_3$ near saturation,
which in a certain range of parameters features six energy minima at $\pi/3$ rotation-symmetric momenta inside the Brillouin zone.
Besides being pedagogical for our study, this model has been proposed for several materials,
such as NiBr$_2$ (Ref.~\onlinecite{nibr2}) and NiGa$_2$S$_4$ (Ref.~\onlinecite{niga2s4}), {both with spin $S=1$}.
Moreover, a recent classical Monte-Carlo study, for a specific choice of exchange couplings,
reported the appearance of an exotic triple-$q$ state,
which is accompanied by skyrmion lattice, at finite temperature in intermediate applied magnetic
field.\cite{okubo2012multiple}
%Here our investigation is somewhat complementary since, although it is restricted to zero temperature
%and high magnetic field, the values of the exchange couplings can be chosen at will.

In this paper, to study possible magnetic phases of the $S=1/2$
triangular-lattice $J_1$-$J_2$-$J_3$ antiferromagnet
near the saturation field, we use the dilute Bose gas theory.
In Sec.~\ref{becdeg}, we write down the general form of ground-state energy \`a la Ginzburg-Landau
for six complex order parameters corresponding to the condensed magnon modes in the dilute limit. {We stress that the same type of effective theory can arise from very different microscopic Hamiltonians. A recent attractive example is given by the spin-dimer compound Ba$_3$Mn$_2$O$_8$, which features magnetic triangular lattices
with non-trivial stacking and interlayer exchange couplings.\cite{kamiya2013magnetic}}
All of the effective coupling constants in this energy {functional} can be calculated from the microscopic model
in the dilute Bose gas approximation.\cite{beliaev1958application,* beliaev1958energy}
This will be done in two ways. First we consider layered systems with a finite, eventually very small,
interlayer coupling;
while the relevant physics still comes from the triangular lattice,
the three-dimensionality naturally protects the calculation from infra-red singularities.
Besides, we take a purely two-dimensional
(2D) approach, in which an infra-red momentum cutoff is introduced as a regularization.
It should be however noted that the latter approach requires the {\it assumption} of a stable low-density
single-magnon Bose gas.
In the present model we find that various instabilities that may affect the existence
of dilute single-magnon gas can not be captured in this approach.

Minimization of the ground-state energy leads to the phase diagrams  of Figs.~\ref{pd13} and \ref{pd12},
which are the main results of this paper.
In particular, besides the well-known spiral and fan phases, we find in quite extended regions
a new phase (``01'' phase) with a striped chiral order and  new multi-ferroic properties,
as described in Sec.~\ref{sec:01}. Also, we show that the presence of ferromagnetic exchange
interactions can sometimes induce an effective attraction between magnon modes, causing an
instability of the dilute magnon gas for weak interlayer coupling regime.
In this situation a field-induced first-order phase transition
(magnetization jump) or a transition to a different quantum phase (not described by a single-magnon BEC)
is typically expected.\cite{ueda2011nematic}
It has been discussed that
ferromagnetic interactions sometimes induce formation of two-magnon bound states, which give rise
to spin nematic ordering.\cite{shannon2006nematic}
In our model, we
indeed find that two-magnon bound states are more stable than single magnons
in a certain parameter region inside of the ``phase separation" region.
{We also applied exact diagonalization analysis of finite-size systems in this
parameter region,
which indicates a small magnetization jump at the saturation field and a tendency toward spin nematic ordering
below the jump.
}

Comparison between purely two-dimensional analysis and quasi-two-dimensional analysis with weak
interlayer coupling also reveals that the shape of a phase boundary can have strong interlayer  coupling ($J_0$)
dependence in the weak $J_0$ limit. The ``01'' phase in Fig.~\ref{pd13} extends to the weak $J_0$ regime,
such as
$J_0\sim10^{-4}$, but purely two-dimensional analysis concludes that this phase cannot appear
in the two-dimensional system near saturation.
We discuss that this singularity comes from the logarithmic correction
of the effective coupling $\Gamma\sim \alpha/(\log|J_0|)+{\cal O}(1/(\log|J_0|)^2)$
and the phase boundary between the ``01'' phase and fan phase presumably has a logarithmic
singularity, going rapidly down to the $J_3=1/4$ point in the $|J_0|\rightarrow 0$ limit.

The paper is organized as follows:
In Sec.~\ref{model} we  briefly describe the model and degeneracy in
the single-magnon energy dispersion at saturation field.
In Sec.~\ref{becdeg} we discuss the dilute Bose gas theory for describing
magnon condensation at multiple momenta, explaining how effective couplings are
calculated from the microscopic model.
In Sec.~\ref{sec:pd_interlayer}  we present results of phase diagrams and
characteristic of each phase.
In Sec.~\ref{sec::concl} we conclude with a summary and discussions.

% In Sec.~\ref{model} we introduce the model and identify the regions of the parameter space that will be of interest to our study. In Sec.~\ref{becdeg} we write down a general effective ground state energy \`a la Ginzburg-Landau

\section{Model} \label{model}

We consider the spin $S=1/2$ $J_1$-$J_2$-$J_3$ model on the triangular lattice in applied magnetic field
at zero temperature
and, including an interlayer coupling, we also consider the model on the hexagonal lattice.
The Hamiltonian reads
\begin{align}
{\cal H}=& J_1 \sum_{\bra i,j \ket}  \bs_i \cdot \bs_j + J_2 \sum_{\bra i,j \ket_{\rm 2nd}}  \bs_i \cdot \bs_j
+J_3 \sum_{\bra i,j \ket_{\rm 3rd}} \bs_i \cdot \bs_j \nonumber\\
& + J_0 \sum_{\bra i,j \ket_\perp}  \bs_i \cdot \bs_j  -H \sum_i S_i^z,
\label{micr}
\end{align}
where $\bra i,j \ket$ counts nearest neighbor bonds, $\bra i,j \ket_{\rm 2nd}$
counts next-nearest neighbor bonds, and $\bra i,j \ket_{\rm 3rd}$ counts 3rd-nearest neighbor bonds
on the triangular-lattice layers.
The $J_0$ term represents the nearest-neighbor (NN) coupling between adjacent layers.
In this paper, we focus on ferromagnetic (negative) $J_1$,
fixing $J_1=-1$ without loss of generality.

The saturation field is defined as the value of the applied magnetic field at which all spins are polarized. Slightly below the saturation field the magnetic excitations  are  interacting hard-core bosons (magnons). The bosonic vacuum corresponds to the fully polarized state.  Using the hard-core boson map \cite{1956PThPh..16..569M} of spin 1/2 operators ($ S^-_i = a_i^\dag$, $ S^z_i=1/2 - a_i^\dag a_i $) the Hamiltonian in Fourier space becomes, modulo constant terms,
\begin{align}
{\cal H}= \sum_{\bk} [\epsilon(\bk) -\mu]\, a^\dag_{\bk }a_{\bk}
+ \frac{1}{2N} \sum_{\bk,\bk',\bq}  V(\bq ) \, a^{\dag}_{\bk+\bq} a^{\dag}_{\bk'-\bq} a_{\bk' } a_{\bk},
\label{boseham}
\end{align}
where $N$ is the number of lattice sites and
\begin{align}
\epsilon (\bk)  =& J_1 \,\nu(\bk) +J_2\, \gamma(\bk)+J_3\, \sigma(\bk) \non
& +J_0 \cos k_z +|J_0|, \\
\nu(\bk) =& \sum_{i=0}^2 \cos {\bf a}_{2i}\cdot \bk, \\
\gamma(\bk) =& \sum_{i=0}^2 \cos {\bf b}_{2i} \cdot \bk,\\
\sigma(\bk) =& \sum_{i=0}^2 \cos {\bf c}_{2i} \cdot \bk,\\
V(\bq) =&  2(\epsilon(\bq) -|J_0| +U),\\
\mu =& 3 (J_1+J_2+J_3) +|J_0| - H.
\end{align}
$\{ {\bf a}_i\}_{i=0,\ldots,5}$, $\{ {\bf b}_i\}_{i=0,\ldots,5}$, and $\{ {\bf c}_i\}_{i=0,\ldots,5}$ are,
respectively, the NN, 2nd-NN, and 3rd-NN lattice vectors of the triangular lattice.
$U$ represents a repulsive on-site interaction,
which will be eventually sent to infinity
to implement the hard-core condition. The saturation field is given by
$H_c = 3 (J_1+J_2+J_3) +|J_0| - \epsilon_{\rm min}$, where $\emin=\min_{\bk} \epsilon(\bk)$.

\begin{figure}[tb]
\begin{center}
\includegraphics[scale=0.5]{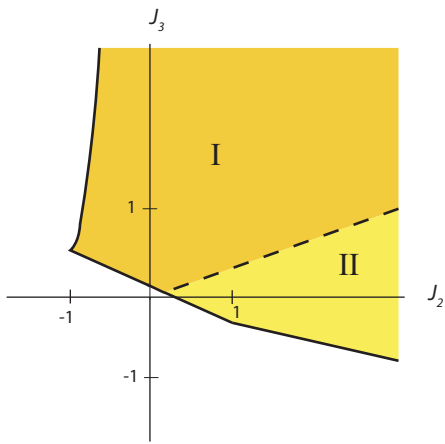}
\end{center}
\caption{Regions of interest (I and II) in the $J_2$-$J_3$ plane {with $J_1=-1$}.
For the analytic expression of the three curves delimiting the colored area we refer to Ref.~\onlinecite{rastelli1979non}. Region I and II are separated by the
line $J_2=2J_3$.}
\label{regions}
\end{figure}

The single-magnon energy minima have qualitatively different structure depending on the value of the exchange couplings.
For ferromagnetic $J_1$ ($J_1=-1$),
there are two interesting regions in the $J_2$-$J_3$ plane\cite{rastelli1979non} (see Fig.~\ref{regions}), with six degenerate minima at
inequivalent (generically incommensurate) wave-vectors.
In region I, they are  $\bQ_0=(k_0^I,0,0)$ (resp. $\bQ_0=(k_0^I,0,\pi)$) for $J_0<0$ (resp. for $J_0>0$) and all $\pi/3$ rotations thereof around the $k_z$ axis; $k_0^I$ is given by
\begin{equation}
k_0^I=2 \cos^{-1} \left( \frac{2J_3-3J_2 +\sqrt{(3J_2+2J_3)^2 +8J_3 }}{8J_3} \right).
\label{kamp1}
\end{equation}
In region II, we instead define $\bQ_0=(0,k_0^{II},0)$ (resp. $\bQ_0=(0,k_0^{II},\pi)$) for $J_0<0$ (resp. for $J_0>0$), with
\begin{equation}
%\frac{J_1}{J_3}= - \frac{2(\sin k +\sin 2k)}{\sin k + \sin \frac{k}{2}}
k_0^{II}=\frac{2}{\sqrt{3}}\cos^{-1}\left(\frac{1-J_2}{2(J_2+2J_3)}\right),
\label{kamp2}
\end{equation}
and the other ones are generated by the same rotational symmetry.

In this paper we are interested in these two areas,
where the system can possibly host new multiple-$q$ phases. In particular, we concentrate on two representative semi-infinite lines, namely i) $J_2=0, J_3>1/4$ for region I and ii) $J_3=0, J_2>1/3$ for region II,
which correspond to the $J_1$-$J_2$ model and the $J_1$-$J_3$ model respectively.
We do not expect qualitative differences for other choices of the parameters within the two regions.

% Focusing on positive (antiferromagnetic) $J_3$, there are four cases: i) for $J_1 \le -4 J_3$ there is only one minimum at  $\bk=(0,0)$; ii) for $-4 J_3 < J_1 < 0$ there are six degenerate minima at (generically incommensurate) wave-vectors $\bk = \pm (k_0,0), \pm  (k_0/2,\sqrt{3}k_0/2), \pm (k_0/2,-\sqrt{3}k_0/2)$, with $k_0=|\bk|$  given by
%\begin{equation}
%%\frac{J_1}{J_3}= - \frac{2(\sin k +\sin 2k)}{\sin k + \sin \frac{k}{2}}
%k_0=\frac{2}{\sqrt{3}}\cos^{-1}\left(\frac{-J_1-J_2}{2(J_2+2J_3)}\right)
%\label{kamp2}
%\end{equation}
%iii) for $J_1=0$ there are four independent sublattices and the problem is reduced to the Heisenberg model on each of them; iv) for $J_1>0$ there are two inequivalent minima at $\bk=\pm (4\pi/3,0)$ and the picture is quite similar to  that of the Heisenberg model.
%
%In this paper we are interested in case ii) since in this range of parameters the system can possibly host new multiple-$q$ phases.
\begin{figure}[tbh]
\begin{center}
\includegraphics[scale=0.5]{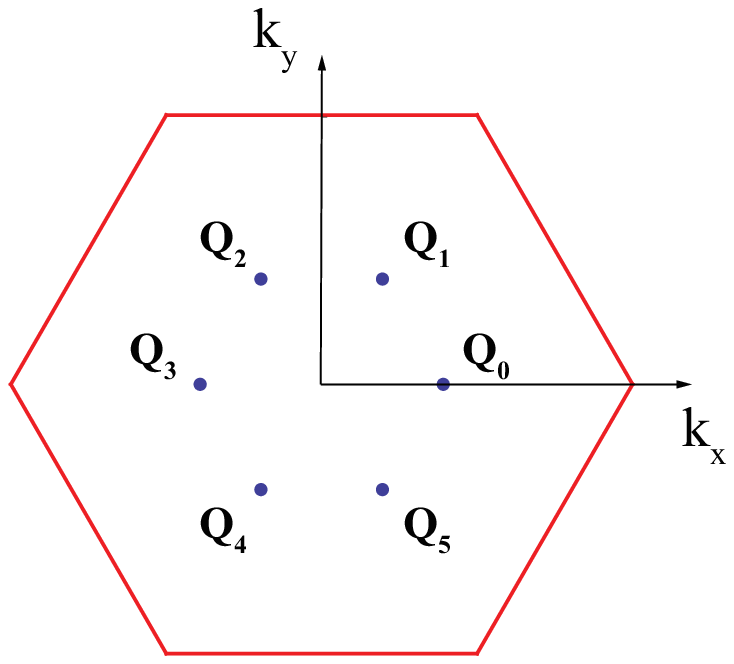}
\includegraphics[scale=0.6]{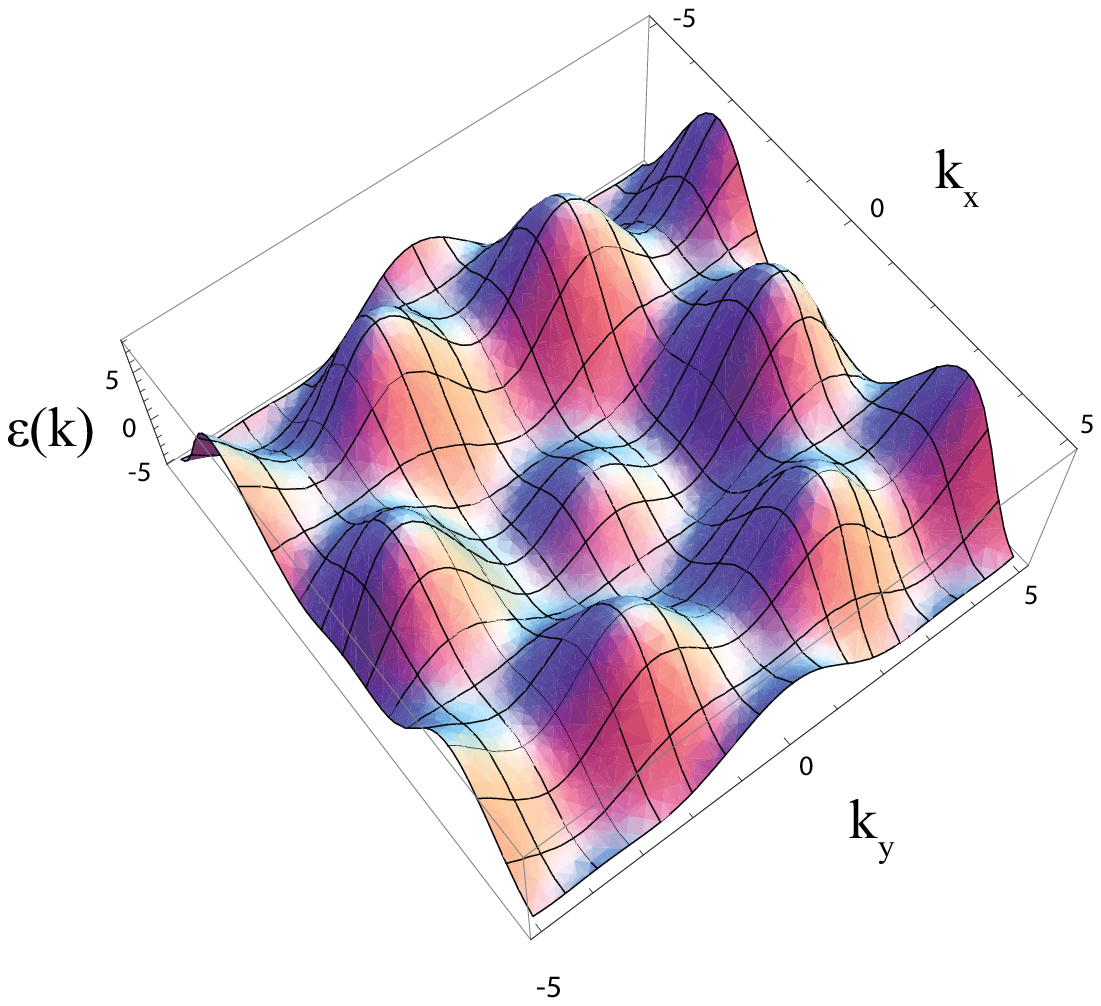}
\end{center}
\caption{Section of the Brillouin zone with the single magnon minima and
single-magnon energy dispersion $\epsilon(\bk)$ in region I
of the $J_2$-$J_3$ parameter space;
in region II they appear  rotated by $\pi/2$.}
\label{caseb}
\end{figure}

\section{Magnon condensation with finite degeneracy} \label{becdeg}

%Without loss of generality we fix $J_1=-1$.
% In the region $J_3 > 0.25$ (case ii)

For applied magnetic field $H$ above the saturation field $H_c$, or in other words for $\mu < \emin$, all spins are aligned along the direction of the field, which corresponds to the absence of magnons. When $H$ is tuned slightly below $H_c$ we expect a dilute gas of  magnons, most of which occupy the lowest energy states.

\subsection{Ground-state energy in the dilute limit}
The six inequivalent single-magnon minima, denoted $\{\bQ_i\}_{i=0,\ldots,5}$, are arranged for region I as in Fig.~\ref{caseb}, where we depict the appropriate section of the Brillouin zone (for region II they are rotated by $\pi/2$).
%If $R_{\pi/3}$ is the $SO(2)$ rotation of angle $\pi/3$ and $\bQ_0=(k,0)$, with $k$ given by Eq.~(\ref{kamp}), then $\bQ_i=(R_{\pi/3})^i \bQ_0$.
We introduce the (complex) order parameters $\bra a_{\bQ_i}\ket = \sqrt{N} \psi_{\bQ_i}$
($i=0,\dots,5$)
referring to particles condensed at the six different wave-vectors $\bQ_i$.
In the dilute limit the ground-state energy per site can be written, by exploiting the
symmetries of the system (six-fold rotation and mirror symmetries), as
\begin{align}
\frac{\cal E}{N} =& \, (\emin -\mu) \sum_{i=0}^5 |\psi_{\bQ_i}|^2+ \frac{1}{2} \Gamma^{(1)} \sum_{i=0}^5 |\psi_{\bQ_i}|^4 \non
& + \sum_{i=0}^5 \sum_{j=1}^2 \Gamma^{(2)}_{j} |\psi_{\bQ_i}|^2 |\psi_{\bQ_{i+j}}|^2 \non
& + \sum_{i=0}^2 \sum_{j=0}^2 \Gamma^{(3)}_{j} \psi^*_{\bQ_{i}}  \psi^*_{-\bQ_{i}} \psi_{\bQ_{i+j}} \psi_{-\bQ_{i+j}}
\label{gse}
 \end{align}
and higher orders in the condensate amplitudes can be neglected.
The coefficients $\Gamma^{(1)}$ and
$\Gamma^{(i)}_j$ are the effective vertices, namely renormalized four-point functions, describing the interaction between condensed particles, that in the dilute regime can be determined by a full quantum mechanical calculation as first shown by Beliaev.\cite{beliaev1958application,*beliaev1958energy}
{The energy} ${\cal E}/{N}$ is clearly real-valued, even though
{not all of} the quartic terms are density-density type;
in particular, the last term
of Eq.~(\ref{gse}) depends on the relative phases of the condensates.
This is a peculiarity of our theory, originating essentially from the presence of
frustrated non-NN
exchange.\footnote{In three-dimensional lattices such as bcc and fcc lattices there can be similar
phase-dependent quartic terms induced by umklapp scattering. }
%compared to the previous theories of magnon condensation with only one\cite{batyev1984antiferromagnet} or two\cite{nikuni1995hexagonal} order parameters.
Note that, while there is only
one global $U(1)$ symmetry in the original spin model [Eq.~(\ref{micr})],
the low-energy effective theory in the dilute limit
enjoys an additional emergent symmetry, namely the product  of three ``chiral'' symmetries $U(1)_j$ ($j=0,1,2$)
acting as $(\psi_{\bQ_j} ,\psi_{-\bQ_j} )\to (e^{i\alpha_j}\psi_{\bQ_j} ,e^{-i\alpha_j}\psi_{-\bQ_j} )$.
\footnote{Without the phase-dependent {last term of Eq.~(\ref{gse}), the symmetry}
would be enhanced to $U(1)^6$, the natural symmetry of a six-species gas with generic density-density interactions.}

In order to find the effective couplings $\Gamma^{(n)}$ in Eq.~(\ref{gse})
we must calculate the renormalized scattering amplitude\footnote{In general
the four-point function $\Gamma(\bq;\bk,\bk';E)$ in our parametrization  is function of exchanged momentum, incoming momenta and total energy, but here we do not indicate the last one unless needed.}
$\Gamma(\bq;\bk,\bk')$ at low density (many-body $T$-matrix)  for initial momenta $\bk,\bk'\in\{\bQ_i\}$.
The effective couplings are given by the following combinations:
\begin{align}
\Gamma^{(1)} = &\Gamma(\b0;\bQ_0,\bQ_0), \label{casebi} \\
 \Gamma^{(2)}_1 = &\Gamma(\b0;\bQ_0,\bQ_1) + \Gamma(\bQ_1-\bQ_0;\bQ_0,\bQ_1), \\ %= \Gamma(\b0;\bQ_0,\bQ_1) + \Gamma(\bQ_2;\bQ_0,\bQ_1), \\
 \Gamma^{(2)}_2 =& \Gamma(\b0;\bQ_0,\bQ_2) + \Gamma(\bQ_2-\bQ_0;\bQ_0,\bQ_2), \\  %= \Gamma(\b0;\bQ_0,\bQ_2) + \Gamma((-\frac{3}{2}k,\frac{\sqrt{3}}{2}k );\bQ_0,\bQ_2), \\
 \Gamma^{(3)}_0 = & \Gamma(\b0;\bQ_0,\bQ_3 ) + \Gamma(\bQ_3-\bQ_0;\bQ_0,\bQ_3), \\
 \Gamma^{(3)}_1 =&\Gamma(\bQ_1-\bQ_0;\bQ_0,\bQ_3) \non
&\qquad + \Gamma(-\bQ_1-\bQ_0;\bQ_0,\bQ_3), \\
\Gamma^{(3)}_2= & \Gamma(\bQ_2-\bQ_0;\bQ_0,\bQ_3) \non
&\qquad + \Gamma(-\bQ_2-\bQ_0;\bQ_0,\bQ_3) = \Gamma^{(3)}_1. \label{casebf}
 %= \Gamma(\bQ_2;\bQ_0,-\bQ_0) + \Gamma((-\frac{3}{2}k,-\frac{\sqrt{3}}{2}k );\bQ_0,-\bQ_0) \\
% & \Gamma^{(3)}_2 = \Gamma(\bQ_2-\bQ_0;\bQ_0,-\bQ_0) + \Gamma(-\bQ_2-\bQ_0;\bQ_0,-\bQ_0)  =  \Gamma((-\frac{3}{2}k,\frac{\sqrt{3}}{2}k );\bQ_0,-\bQ_0)+ \Gamma(-\bQ_1;\bQ_0,-\bQ_0)
\end{align}
The strategy for the calculation is presented in the following Section~\ref{sec:bseq}.
%** Description of various possible phases? **

\subsection{Bethe-Salpeter equation} \label{sec:bseq}
\begin{figure}[bt]
\includegraphics[scale=0.5]{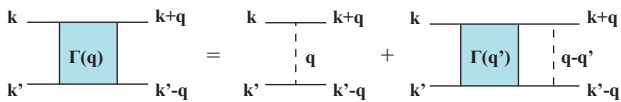}
\caption{Ladder diagram included in Bethe-Salpeter equation (\ref{bseq}).
The filled squares and the dashed lines represent the full and the bare interaction respectively.}
\label{ladder}
\end{figure}

In the dilute limit  $\Gamma(\bq;\bk,\bk')$ satisfies the Bethe-Salpeter equation for the ladder approximation, which  reads
\begin{widetext}
\begin{align}
 \Gamma(\bq;\bk,\bk') =  V(\bq)  - \frac{1}{N} \sum_{\substack{\bq' \in {\rm BZ}\\ \bq':\epsilon-\emin>\mu} } \frac{V(\bq-\bq')}{\epsilon(\bk+\bq')+\epsilon(\bk'-\bq')-2\emin -E} \Gamma(\bq';\bk,\bk').
\label{bseq}
\end{align}
\end{widetext}
We keep the total energy $E$ (measured from the minimum of the free two-particle spectrum) generically different from its on-shell value $E=0$ and an infrared cutoff for reasons that will become clear in the following. This equation is diagrammatically depicted in Fig.~\ref{ladder}.

%\begin{widetext}
%
%\begin{figure}[htb]
%\begin{center}
%\begin{fmffile}{fgraphs}
%\begin{align*}
%\parbox{140pt}{
%\fmfframe(50,0)(50,0){
%\begin{fmfgraph*}(50,50)
%\fmfleft{i1,i2}
%\fmfright{o1,o2}
%\fmflabel{$\bk$}{i1}
%\fmflabel{$\bk'$}{o1}
%\fmflabel{$\bk+\bq$}{i2}
%\fmflabel{$\bk'-\bq$}{o2}
%\fmf{fermion}{i1,v1,i2}
%\fmf{fermion}{o1,v1,o2}
%\fmfblob{10}{v1}
%\end{fmfgraph*}
%}
%}
%=
%\parbox{140pt}{
%\fmfframe(40,0)(50,0){
%\begin{fmfgraph*}(50,50)
%\fmfleft{i1,i2}
%\fmfright{o1,o2}
%\fmflabel{$\bk$}{i1}
%\fmflabel{$\bk'$}{o1}
%\fmflabel{$\bk+\bq$}{i2}
%\fmflabel{$\bk'-\bq$}{o2}
%\fmf{fermion}{i1,v1,i2}
%\fmf{fermion}{o1,v1,o2}
%\fmfv{decor.shape=circle,decor.filled=0,size=10}{v1}
%\end{fmfgraph*}
%}
%}
%+
%\parbox{130pt}{
%\fmfframe(30,0)(30,0){
%\begin{fmfgraph*}(80,100)
%\fmfleft{i1,i2}
%\fmfright{o1,o2}
%\fmflabel{$\bk$}{i1}
%\fmflabel{$\bk'$}{o1}
%\fmflabel{$\bk+\bq$}{i2}
%\fmflabel{$\bk'-\bq$}{o2}
%\fmf{fermion}{i1,v1}
%\fmf{fermion}{o1,v1}
%\fmf{fermion}{v2,o2}
%\fmf{fermion}{v2,i2}
%\fmf{fermion,label=$k+q'$,left}{v1,v2}
%\fmf{fermion,label=$k'-q'$,right}{v1,v2}
%\fmfblob{10}{v1}
%\fmfv{decor.shape=circle,decor.filled=0,size=10}{v2}
%\end{fmfgraph*}
%}
%}
%\end{align*}
%\caption{Bethe-Salpeter equation. The shaded blobs and the empty circles represent the full and the bare interaction respectively.}
%\label{ladder}
%\end{fmffile}
%\end{center}
%\end{figure}
%\end{widetext}

The ladder approximation includes all multiple scattering of two particles; processes involving more than two particles are indeed suppressed at low density.\cite{abrikosov1975methods,schick1971two}
This actually amounts to approximating $\Gamma(\bq;\bk,\bk')$
with the renormalized scattering amplitude for two particles in the vacuum (two-body $T$ matrix), namely at $\mu-\emin=0$.   While in three dimensions
this gives a finite result that is correct also at low but non-vanishing density  up to small correction
of order $\mu-\emin$, it is well-known that the two-body scattering amplitude  vanishes logarithmically
with lowering the density
in two dimensions\cite{fisher1988dilute}, due to the non-integrable singularities in the kernel of Eq.~(\ref{bseq}).
Thus finite density (many-body) effects become important. In fact, we  expect that at energies lower than $\mu-\emin$ the magnon dispersion is modified \`a la Bogoliubov and becomes linear.  In the calculation  it is therefore required to cutoff the integration in the neighborhoods where $\epsilon(\bk+\bq'),\epsilon(\bk'-\bq')\lesssim \mu$.
This is the meaning of the cutoff introduced in  Eq.~(\ref{bseq}).
Whereas it is possible to work directly in momentum space, we choose the more convenient treatment of Refs.~\onlinecite{morgan2002off,*lee2002energy} (see also Ref.~\onlinecite{morgan2000gapless}), where it is shown that calculating Eq.~(\ref{bseq}) at negative energy $E=-C(\mu-\emin)$ ($C$ a numerical constant of order 1) {without momentum cutoff} yields an equivalent result at leading order in $\mu-\emin$.
%(see Ref.~\onlinecite{kolezhuk2012frustrated} for a recent application to a frustrated spin system near saturation).
This procedure will be used in Section~\ref{sec:pure2d}, while in the following Section~\ref{sec:pd_interlayer}
we take a different approach, namely we consider the system with
a non-vanishing interlayer coupling, thus avoiding the subtleties appearing in two dimensions.

%
%This is clearly an artifact of approximating the single-magnon dipersion with the free one, without taking into account the interaction with the medium (condensate). On the other hand, we actually expect that at energies lower than $\mu-\emin$ the magnon dispersion is modified \`a la Bogoliubov and becomes linear, thus smoothing out the pathological singularities. In the calculation of Eq.~(\ref{ladder}) it is therefore required to cutoff the integration in the neighborhoods where $|\epsilon(\bk+\bq')+\epsilon(\bk'-\bq')-2\emin|\lesssim \mu-\emin $. Whereas this is doable directly in momentum space, we choose the more convenient treatment of Refs.~\onlinecite{morgan2002off,*lee2002energy} (see also Ref.~\onlinecite{morgan2000gapless}), where it is shown that calculating Eq.~\ref{bseq} at negative energy $E=-C(\mu-\emin)$ ($C$ numerical constant of order 1) yields an equivalent result (see \onlinecite{kolezhuk2012frustrated} for a recent application to a frustrated spin system near saturation). This procedure will be used in the following section, while in Section~\ref{interlayer} we take a different approach, namely we introduce a non-vanishing interlayer coupling.

\section{Phase diagrams: Quasi-2D systems with interlayer coupling} \label{sec:pd_interlayer}

We now consider layered systems with non-vanishing interlayer NN exchange coupling $J_0$.
Whereas the interesting physics  is essentially delivered by the triangular lattice planes,
the sign of $J_0$ determines the  relative ordering of two adjacent planes.

%Let us start with ferromagnetic $J_0<0$.  The bosonic Hamiltonian has the same form of Eq.~(\ref{boseham}) with
%\begin{align}
%&\epsilon (\bk)  = J_1 \,\nu(\bk) +J_3\, \sigma(\bk) +J_0 ( \cos k_z-1) \\
%&V(\bq) = 2( J_1 \,\nu(\bq) +J_3\, \sigma(\bq) +J_0  \cos k_z +U)\\
%&\mu = 3 (J_1+J_3) -J_0 - H
%\end{align}
%where the extension of the lattice vector to 3D is obvious and we have added a constant to the energy for convenience, such that it takes the same minimal values $\epsilon_{min}$ as in the 2D case.
The solution of Eq.~(\ref{bseq}) can be obtained by expanding
in lattice harmonics, that is by taking the  Ansatz
\begin{align}
\Gamma(\bq)= & \bra \Gamma \ket +\sum_{i=0}^2 \{ J_1 A_i \cos {\bf a}_{2i}\cdot \bq + J_2 B_i \cos {\bf b}_{2i}\cdot \bq \non
& +  J_3 C_i \cos 2{\bf a}_{2i}\cdot \bq  \}  +J_0  D \cos {q_z},
\label{ansatz3}
\end{align}
where $\bra \Gamma \ket=(1/N)\sum_q \Gamma(\bq)$.
The calculation of the effective coupling is detailed in Appendix~\ref{bscalc}; in this case we do not need any cutoff procedure since all integrals are finite.

%The antiferromagnetic case $J_0>0$ is analogous. For similar reasons as above we prefer to change the constant in the definition of $\epsilon(\bk)$,
%\begin{align}
%\epsilon (\bk)  = J_1 \,\nu(\bk) +J_3\, \sigma(\bk) +J_0 ( \cos k_z+1)
%\end{align}
%The minima are $\bQ_0=(k,0,\pi)$, etc..

\begin{figure}[tb]
\begin{center}
\includegraphics[width=0.44\textwidth]{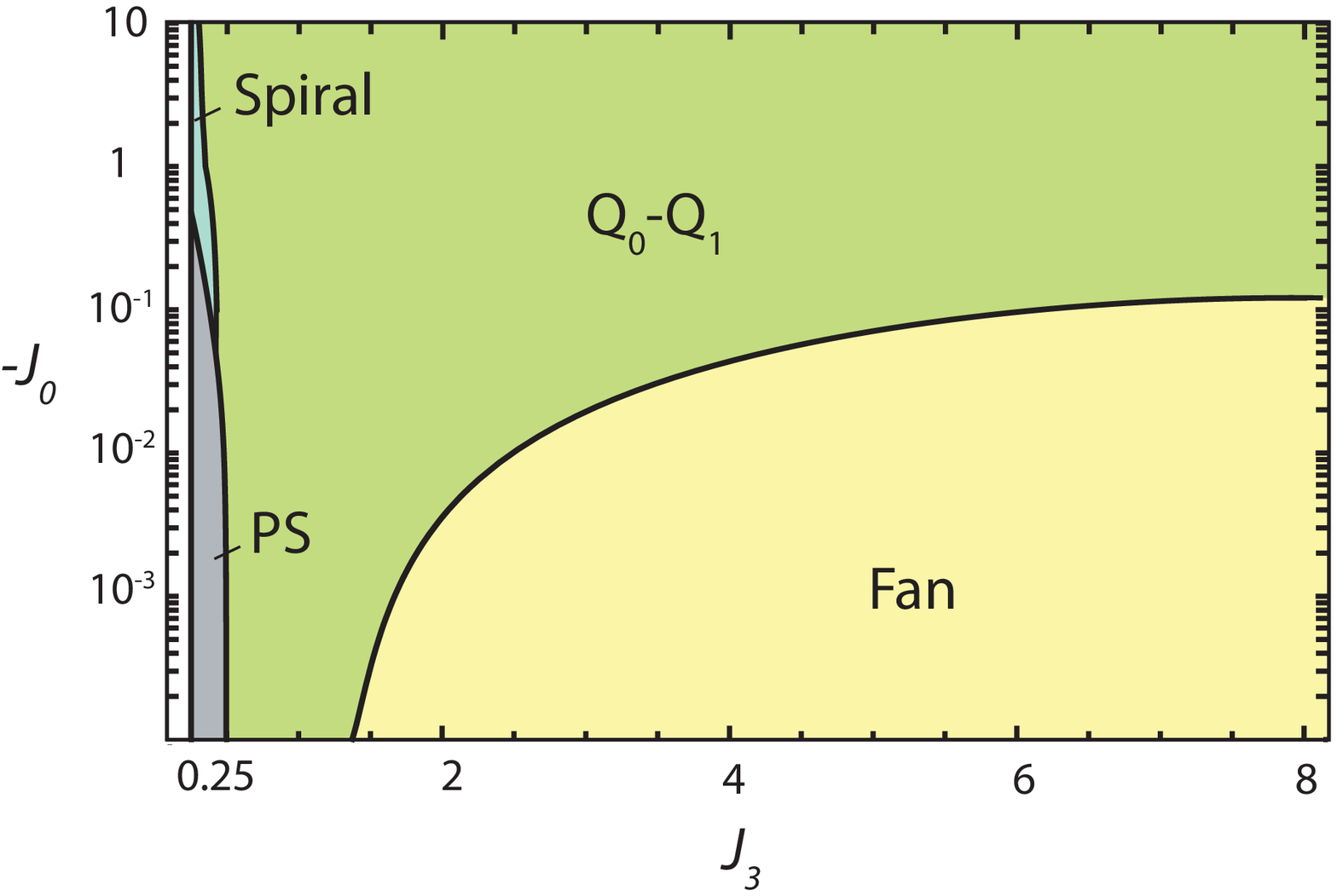}
\includegraphics[width=0.44\textwidth]{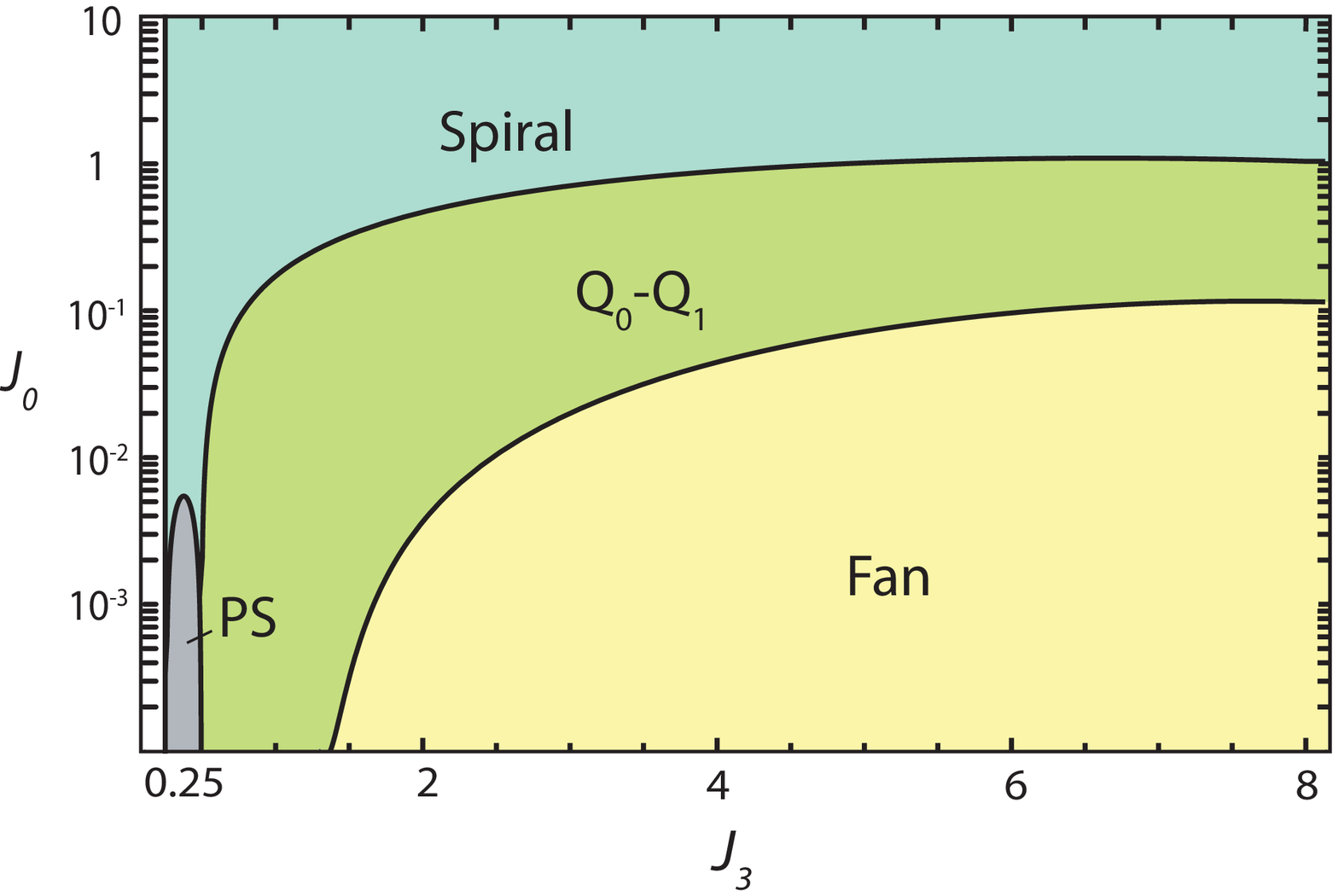}
\end{center}
\caption{Phase diagrams of the $J_1$-$J_3$ model
for ferromagnetic (upper) and antiferromagnetic (lower) interlayer coupling $J_0$,
{with $J_1=-1$}.
{PS denotes the regions with phase separation (see Sec.~\ref{sec:PS})
and $Q_0$-$Q_1$ denotes ``01" phase (see Sec.~\ref{sec:01}).}
Note the logarithmic scale of the vertical axis.}
\label{pd13}
\end{figure}

\begin{figure}[!tb]
\begin{center}
\includegraphics[width=0.44\textwidth]{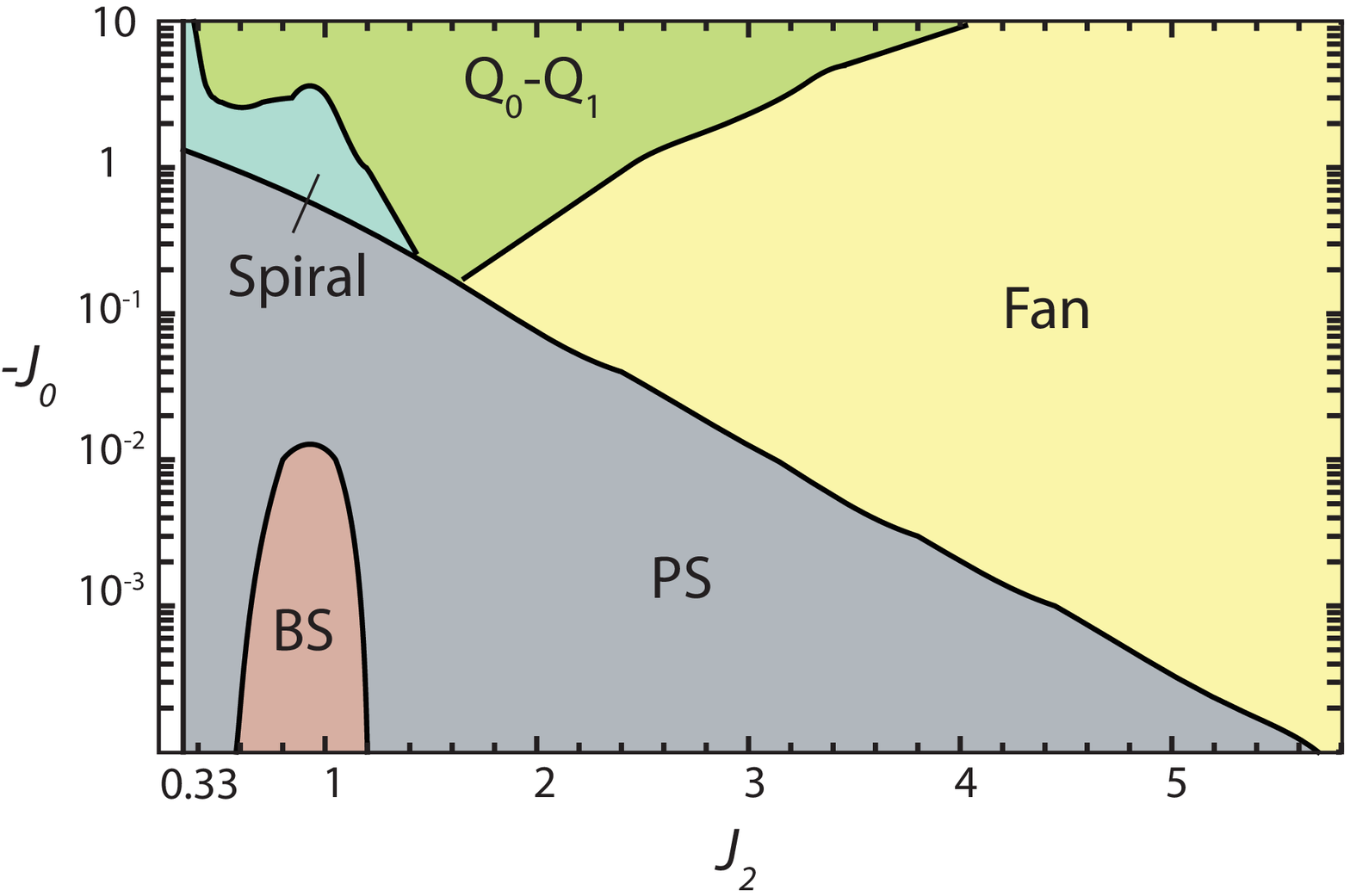}
\includegraphics[width=0.44\textwidth]{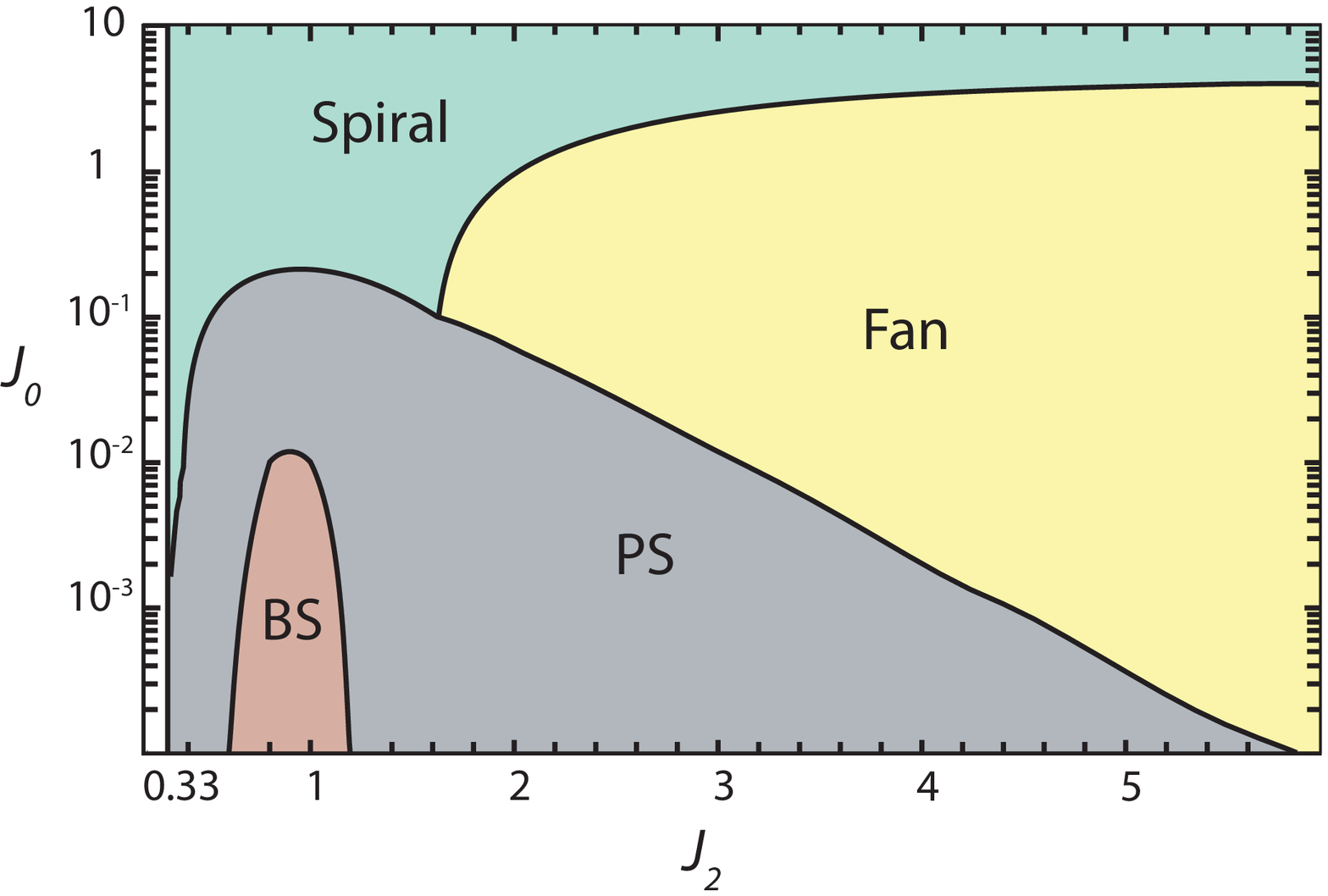}
\end{center}
\caption{Phase diagrams of the $J_1$-$J_2$ model
for ferromagnetic (upper) and antiferromagnetic (lower)
interlayer coupling $J_0$, {with $J_1=-1$}.
%PS denotes the regions with phase separation (see Sec.~\ref{sec:PS}),
%$Q_0$-$Q_1$ denotes ``01" phase (see Sec.~\ref{sec:01}), and
BS denotes the regions where two-magnon
bound states have lower energy.
}
\label{pd12}
\end{figure}

By plugging the result into and minimizing Eq.~(\ref{gse}) we obtain
the phase diagrams of the $J_1$-$J_3$ and $J_1$-$J_2$ models,
which are shown in Figs.~\ref{pd13}
and \ref{pd12} respectively.
It is interesting to note that in the classical limit, these models always have  the spiral  state
in their ground state manifold in the present parameter
space. However, this phase disappears in most cases due to quantum effects and is replaced with other
new quantum phases.

Below we describe the characteristics of the different regions
composing the phase diagrams.

%*** Some comments are in order. For any value of $J_0$, in the large $J_3$ limit the ground state is the coplanar state
%%Eq.~(\ref{coplanar})
%since $\Gamma^{(3)}_0$ is the smallest coupling. This is in accordance with the expectations from \onlinecite{nikuni1995hexagonal}: in fact for $J_3\gg |J_1|$ our system should behave similarly to a NN triangular antiferromagnet (with supercell made of four original unit cells), where the coplanar state is stabilized for all $J_0<0$. As $J_3$ decreases the system enters another two-$q$ phase characterized by adjacent wave-vectors $\bQ_0,\bQ_1$ (or pairwise $\pi/3$ rotations thereof), that we denote simply ``01-phase''. The critical value $J_{3c}$ depends on $J_0$, namely it decreases for decreasing $|J_0|$. ***
%%approximate values are $J_{3c}=20, 6, 2.5,1.7$ for $J_0=-1,-0.1,-0.01,-0.001$ respectively.

%For $-0.1 \lesssim J_0 <0$ and small $J_3$ ($J_3\lesssim 0.3$) the two couplings $\Gamma^{(1)},\Gamma^{(3)}_0$ become slightly negative, which indicates a possible instability to be further investigated.
%Fig.~\ref{fig:all3dzoom} shows a detailed view of these regions.  Note that the calculation is quite noisy in the vicinity of the critical point  $J3=0.25$ (where the $\bQ_i$'s merge into (0,0,0) and the ground state is becomes the ferromagnetic state). It is reasonable to consider that the couplings actually vanish for $J_3=0.25$.

%We believe that this effect is due to the ferromagnetic $J_0$, since it doesn't appear in the 2D limit $J_0\to 0$. In order to support this argument we study the antiferromagnetic case $J_0>0$.

\subsection{Spiral phase} \label{sec:spiral}

Magnon condensation at a single wave-vector, say $\psi_{\bQ_0}=\sqrt{\rho}e^{i\alpha}$ and
$\psi_{\bQ_i}=0$ for $i=1,\cdots,5$, %all the others zero,
yields the so-called spiral phase, whose spin structure is
\begin{align}
&\vev{S^x_j}=  \sqrt{\rho}\cos\left(\bQ_0\cdot \br_j +\alpha \right),   \non
&\vev{S^y_j}= \sqrt{\rho}\sin\left(\bQ_0\cdot \br_j+\alpha\right),  \non
&\vev{S^z_j}= \frac{1}{2}- \rho.
\label{spiral}
\end{align}
The magnon density is given by $ \rho=(\mu-\emin)/\Gamma^{(1)}$.
This phase breaks the $C_6$ rotation symmetry and reflection symmetry,
and is accompanied by a vector chiral order
\begin{align}
(\vev{{\bf S}_{\bf r}} \times \vev{{\bf S}_{{\bf r}+{\bf a}_0}})^z=\rho \sin(\bQ_0\cdot {\bf a}_0).
\end{align}
As noted already in
Ref.~\onlinecite{ueda2009magnon} this phase shows multi-ferroic behaviour due to the spin-current mechanism,\cite{katsura2005spin} which generates an electric polarization  $ {\bf P}_{\bf e}= \eta \, {\bf e} \times (\vev{{\bf S}_{\bf r}} \times \vev{{\bf S}_{{\bf r}+{\bf e}}})$ associated to a bond $\mathbf{e}$ ($\eta$ is a constant). Specifically, for longitudinal magnetic field $\mathbf{H}=(0,0,H)$,
\begin{align}
{\bf P}_{{\bf a}_i} =\eta\, \rho \sin \left( \mathbf{Q}_0 \cdot \mathbf{a}_i \right) (\widehat{\mathbf{a}}_i \times \widehat{\mathbf{H}} ).
\end{align}
Let us note that the out-of-plane component of ${\bf P}_{{\bf a}_i} $ is locally non-zero, but it vanishes in average over a period since it is proportional to $\sin\left( \mathbf{Q}_0 \cdot ( \mathbf{r}+\mathbf{a}_i /2)\right)$. In case of in-plane magnetic field all components vanish in average.

\subsection{Fan phase} \label{sec:coplanar}
In this phase magnons condense simultaneously at two opposite wave-vectors,
{\it e.g.} $\bQ_0$ and $\bQ_3 {\cong} -\bQ_0$ {with the same density}.
We can choose the parametrization $\psi_{\bQ_0}=\sqrt{\rho} e^{i\alpha_1}$ and
$\psi_{\bQ_3}=\sqrt{\rho} e^{i\alpha_2}$ with  $\rho=(\mu-\emin)/(\Gamma^{(1)} +\Gamma^{(3)}_0)$,
and define $\theta=\alpha_1+\alpha_2$, $\phi=\alpha_2-\alpha_1$. The  spin structure is given by
\begin{align}
&\vev{S^x_j}= 2 \sqrt{\rho}\cos\left(\bQ_0\cdot \br_j-\frac{\phi}{2} \right)  \cos\frac{\theta}{2},  \non
&\vev{S^y_j}= 2 \sqrt{\rho}\cos\left(\bQ_0\cdot \br_j-\frac{\phi}{2} \right)  \sin\frac{\theta}{2},  \non
&\vev{S^z_j}= \frac{1}{2}-4 \rho \cos^2\left(\bQ_0\cdot \br_j-\frac{\phi}{2} \right).
\label{coplanar}
\end{align}
This state has a coplanar spin structure;
The spins oscillate within a fixed plane parallel to the $z$-axis and
identified by the angle $\theta$. %, which is free, as well as $\phi$.
This phase breaks only $C_3$ symmetry and is not accompanied by chiral symmetry breaking.
The vector chirality $\vev{{\bf S}_{\bf r}} \times \vev{{\bf S}_{{\bf r}+{\bf e}}}$ always vanishes
on average and no multi-ferroic property can appear.
%The other phase $\phi$ can in principle be constrained by taking into account subleading correction to Eq.~(\ref{gse}) originating from three-magnon scattering of the type $\Gamma^{3m} [(\psi_{\bQ_0}^*)^3 \psi_{\bQ_3}^3 + c.c ]= 2\rho^3 \cos 3\phi  $, although the calculation is very difficult in practice \cite{nikuni1995hexagonal}.

\subsection{``01" phase} \label{sec:01}
\begin{figure}[tb]
\begin{center}
\includegraphics[scale=0.4]{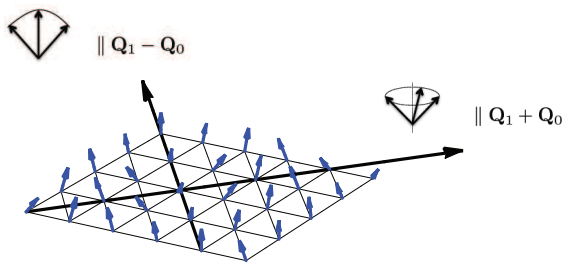}
\end{center}
\caption{Spin structure of the ``01'' phase.}
\label{01phase}
\end{figure}

In the regions denoted by ``$Q_0$-$Q_1$" in Figs.~\ref{pd13} and \ref{pd12},
magnons equally occupy {the lowest-energy} states of two adjacent wave-vectors,
{\it e.g.} $\bQ_0$ and $\bQ_1$.
The spin structure is given by
\begin{widetext}
\begin{align}
&\vev{S^x_j}= 2 \sqrt{\rho}\cos\left(\frac{\bQ_1-\bQ_0}{2}\cdot \br_j+\frac{\phi}{2} \right)
\cos\left(\frac{\bQ_1+\bQ_0}{2} \cdot \br_j+\frac{\theta}{2} \right),   \non
&\vev{S^y_j}= 2 \sqrt{\rho} \cos\left(\frac{\bQ_1-\bQ_0}{2}\cdot \br_j+\frac{\phi}{2} \right)  \sin\left(\frac{\bQ_1+\bQ_0}{2}\cdot \br_j+\frac{\theta}{2} \right), \non
&\vev{S^z_j}= \frac{1}{2}-4 \rho \cos^2 \left(\frac{\bQ_1-\bQ_0}{2}\cdot \br_j+\frac{\phi}{2} \right),
\end{align}
\end{widetext}
where the condensate density is ${\rho}= (\mu-\emin)/(\Gamma^{(1)} +\Gamma^{(2)}_1)$.
This phase somehow interpolates between the spiral and fan phases.
In fact, along the direction of $\bQ_1+\bQ_0$
the spins spiral with pitch vector $(\bQ_1+\bQ_0)/2$,
whereas along the orthogonal direction they oscillate in the fan state (see Fig.~\ref{01phase}).
The ($z$-component of) vector chiral order
exists forming a stripe structure,
\begin{align}
&(\vev{{\bf S}_{\bf r}} \times \vev{{\bf S}_{{\bf r}+{\bf l}}})^z \non
&=
4 \rho \cos^2\left(\frac{\bQ_1-\bQ_0}{2}\cdot \br_j+\frac{\phi}{2} \right)
\sin \left(\frac{\bQ_1+\bQ_0}{2}\cdot {\bf l} \right)
\end{align}
for ${\bf l}\parallel \bQ_1+\bQ_0$, where the stripe of the chiral order
is parallel to the vector $\bQ_1+\bQ_0$.

Recalling the considerations of Sec.~\ref{sec:spiral} we find an induced electric polarization for
longitudinal magnetic field given by
%\begin{widetext}
\begin{align}
{\bf P}_{{\bf a}_i} =4\eta\, \rho \cos\left(\frac{\bQ_1-\bQ_0}{2}\cdot \br_j+\frac{\phi}{2} \right)
\non
\cos\left(\frac{\bQ_1-\bQ_0}{2}\cdot (\br_j+{\bf a}_i)+\frac{\phi}{2} \right) \non
 \sin \left( \frac{\bQ_1+\bQ_0}{2} \cdot \mathbf{a}_i \right) (\widehat{\mathbf{a}}_i \times \widehat{\mathbf{H}} ).
\end{align}
For bonds in the $\mathbf{l}$ direction (${\bf l}\parallel \bQ_1+\bQ_0$),
this expression simplifies to
\begin{align}
{\bf P}_{\bf l} =4\eta\, \rho \cos^2\left(\frac{\bQ_1-\bQ_0}{2}\cdot \br_j+\frac{\phi}{2} \right)
\non
 \sin \left( \frac{\bQ_1+\bQ_0}{2} \cdot \mathbf{l} \right) (\widehat{\mathbf{l}} \times \widehat{\mathbf{H}} ).
\end{align}
%\end{widetext}
%\textcolor[rgb]{1,0,0}{(\textbf{Q}: cosine function also slightly depends on ${\bf a}$?)}
The  main peculiarity compared to the spiral case  is that the amplitude of the polarization  is modulated along one direction, but does not change sign, thus yielding a striped structure with   non-zero net average over a period.

\subsection{Phase separation (PS)}\label{sec:PS}

If  at least one of $\Gamma^{(1)}$, $\Gamma^{(1)}+\Gamma^{(2)}_1$, and
$\Gamma^{(1)}+\Gamma^{(3)}_0$ becomes negative
the system suffers from instabilities as is clear from the runaway behaviour of Eq.~(\ref{gse}).
As discussed recently in Ref.~\onlinecite{ueda2011nematic}, in this situation a state with low density of magnons  can not be stable. The system instead undergoes a field-induced first-order phase transition, featuring phase separation between   the fully polarized state and a low-magnetization state.
Technically, the latter can be stabilized by including higher order terms in the ground state energy Eq.~(\ref{gse})
(e.g. sixth order in the $\psi$'s). By looking at which of the above three combinations of couplings
is (the most) negative, one can argue about the nature of the low-magnetization state.
For instance, if $\Gamma^{(1)}<0$ and all others positive, it is reasonable to expect a low-magnetization spiral state,
etc..

\subsection{Bound states (BS)}

In the $J_1$-$J_2$ model at relatively small interlayer coupling there exist regions
where bound states of two magnons are formed (see Fig.~\ref{pd12}).
This can be inferred from the appearance of a pole singularity in $\Gamma^{(1)}$,
that is essentially a two-particle Green's function at zero frequency.
The presence of a bound state branch in the spectrum below the single-magnon states would
suggest the occurrence of  BEC of bound states and thus a spin nematic
phase.\cite{chubukov91,shannon2006nematic}
We however do not know how the bound states interact and therefore we can not make any statement about
the stability of the spin nematic phase only from this analysis.
If the interaction is attractive, there will be again phase separation.
Moreover, we can not rule out the existence of three magnon bound states or higher.
%\textcolor[rgb]{1,0,0}{
%As discussed in Sec.~\ref{sec::concl}, an
%exact diagonalization study of the triangular-lattice $J_1$-$J_2$ model with 48 spins indeed indicates
%a small magnetization jump at the saturation field and a tendency to the two-magnon pairing below the magnetization
%jump in the vicinity of $J_2=1$, where $J_1=-1$.}
%%from the dependence of the ground state energy on the total magnetization.

\begin{figure}[tb]
\begin{center}
\includegraphics[scale=0.35]{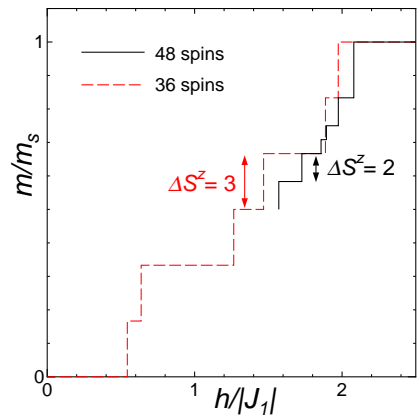}
\end{center}
\caption{{Magnetization process of the $S=1/2$ triangular-lattice
$J_1$-$J_2$ model with $J_1=-1$ and
$J_2=1$ for $N=36$ and 48 spin clusters.
In the smaller size system ($N=36$), the total magnetization always changes by $\Delta S^z=3$,
but, in the larger size system ($N=48$), it has a jump $\Delta S^z=4$ below saturation
and wide steps at even values of $S^z$ below the jump, showing a tendency to the formation of two-magnon bound
states.}}
\label{fig:energy}
\end{figure}

{
To examine the appearance of spin nematic phase we performed
exact diagonalization study of the purely two-dimensional
model $(J_0=0)$ with $N=36$ and 48 spins with the fixed choice of parameters   $J_1=-1$ and $J_2=1$. We used finite-size clusters with high space symmetry ($C_{3v}$)
and under the periodic boundary condition. The magnetization process
is plotted in Fig.\ \ref{fig:energy}. %for $J_2=1$ with $J_1=-1$.
In case of 36 spins, we find a weak signature of formation of three-magnon bound states
from saturation down to low magnetization, which corresponds to the change of total magnetization by three
($\Delta S^z=3$)\cite{momoisn2006}.
However, this seems to be an artefact of a small size system, since for the larger size system ($N=48$),
the magnetization process does not posses this periodicity and, instead, it shows a tendency to the formation of
two-magnon bound states; the lowest energy states in even number $S^z$ sectors have lower energy than in odd $S^z$ sectors,
giving rise to wide steps at even $S^z$ in the magnetization process.
Clearly the stability of this spin nematic phase below the magnetization jump remains to be studied further because finite-size effects can still be strong for $N=48$.}

\subsection{Purely 2D calculation} \label{sec:pure2d}

To compare the quasi-two-dimensional systems with purely two-dimensional systems, we
analyze effective coupling $\Gamma$ in two dimensions, using the procedure
described in Sec.~\ref{sec:bseq}.
%The solution of Eq.~(\ref{bseq}) can be obtained again by expanding in lattice harmonics.
%that is by taking the Ansatz
%\begin{align}
%\Gamma(\bq)= \bra \Gamma \ket +\sum_{i=0}^2 \left\{ J_1 A_i \cos {\bf a}_{2i}\cdot \bq + J_3 B_i \cos 2{\bf a}_{2i}\cdot \bq  \right\}
%\label{ansatz2}
%\end{align}
% as detailed in Appendix~\ref{bscalc}.
At small energy cutoff $E<0$ of order $\mu-\emin$,
the $\Gamma$'s in Eq.~(\ref{casebf}) have a $1/|\log |E||$ expansion  that looks like
\begin{align}
& \Gamma^{(1)}=\frac{\alpha^{(1)}} {\left|\log |E| \right|}  + {\cal O} \left(\frac{1}{\left| \log |E| \right|^2} \right), \\
&\Gamma^{(l)}_m=\frac{\alpha^{(l)}_m } {\left|\log |E| \right|} + {\cal O} \left(\frac{1}{\left| \log |E| \right|^2} \right).
\label{gamm2dexp}
\end{align}
The leading coefficients can be calculated analytically as described in Appendix~\ref{bscalc}.
The results are shown in Fig.~\ref{fig:leading}
for the case of the $J_1$-$J_3$ model,
from which we can see that
\begin{align}
& \alpha^{(1)}=\alpha^{(3)}_0, \qquad \alpha^{(2)}_1=\alpha^{(2)}_2, \qquad \alpha^{(3)}_1=0,\label{alfa1}\\
& 0<\alpha^{(1)}<\alpha^{(2)}_1 \label{alfa2}
\end{align}
for arbitrary $J_3{>1/4}$ ($J_1=-1$ fixed).
An analogous behavior occurs for the $J_1$-$J_2$ model.

\begin{figure}[hb]
\begin{center}
\includegraphics[scale=1.1]{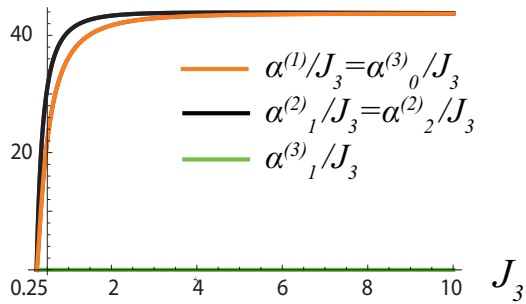}
\caption{Effective couplings (leading coefficients) {of the $J_1$-$J_3$ model}
as a function of $J_3$,
with $J_1=-1$.
%$\alpha^{(1)}={\alpha^{(3)}_0}$
%(orange), $\alpha^{(2)}_1=\alpha^{(2)}_2$ (black),  and  $\alpha^{(3)}_1=0$ (green).
}
\label{fig:leading}
\end{center}
\end{figure}

Thus at leading order the ground-state energy becomes
\begin{widetext}
\begin{align}
\frac{{\cal E}_0}{N}  =&  (\emin -\mu) \sum_{i=0}^5 |\psi_{i}|^2+ \frac{1}{2} \frac{\alpha^{(1)}}{\left|\log |E| \right|}   \left( \sum_{i=0}^5 |\psi_{i}|^4 +2\sum_{j=0}^2 |\psi_{i}|^2 |\psi_{{i+3}}|^2 \right) + \frac{\alpha^{(2)}_1}{\left|\log |E| \right|}  \sum_{i=0}^5 \sum_{j=1}^2 |\psi_{i}|^2 |\psi_{{i+j}}|^2 \non
 =&  (\emin -\mu) \sum_{i=0}^5 |\psi_{i}|^2+  \frac{1}{2} \frac{\alpha^{(1)}}{\left|\log |E| \right|}   \sum_{i=0}^2 \left(  |\psi_{i}|^2 + |\psi_{{i+3}}|^2  \right)^2  \non
& +  \frac{\alpha^{(2)}_1}{\left|\log |E| \right|}
 \left[(|\psi_{0}|^2 + |\psi_{{3}}|^2) (|\psi_{1}|^2 + |\psi_{{4}}|^2)+(|\psi_{0}|^2 + |\psi_{{3}}|^2)(|\psi_{2}|^2 + |\psi_{{5}}|^2) +(|\psi_{1}|^2 + |\psi_{{4}}|^2)(|\psi_{2}|^2 + |\psi_{{5}}|^2)\right],
 \label{gsesymm}
\end{align}
\end{widetext}
which exhibits an emergent $U(2)^3$ symmetry. Namely in the zero density limit,
$\mu \to \emin$  ($E \to 0)$, each $U(1)_i$  is effectively enhanced to $U(2)_i$
whose elements transform the doublet  $(\psi_i, \psi_{i+3})^T$. This is reflected
in the degeneracy of a continuous family of physically distinct ground states, defined by
\begin{align}
&|\psi_{\bQ_i}|^2 + |\psi_{\bQ_{i+3}}|^2 = \frac{\mu-\emin}{\Gamma^{(1)}},\non
&\psi_{\bQ_j}=0 \ \ \ \mbox{if $j\ne i$, $i+3$}
\label{gsmanifold}
\end{align}
for a certain $i$. To the level of approximation of Eq.~\eqref{gsesymm}
the $U(2)_i$ symmetry is then spontaneously broken to $U(1)$ by choosing a ground state
out of the space Eq.~(\ref{gsmanifold}).
The  spiral phase (see Sec.~\ref{sec:spiral}) and coplanar (fan) phase (see Sec.~\ref{sec:coplanar})
are just two states in this ground-state manifold.
This symmetry enhancement is analogous  to that occurring at  low-energy
in a mixture of two species of dilute Bose gases with equal masses in the continuum in two dimensions.\cite{kolezhuk2010stability}
It is worth noting, however, that experiments with cold atoms are typically done at fixed number of particles,
whereas in magnetic systems the chemical potential, which is determined by the applied magnetic field,
can be actually made vanishingly small.
%One might expect that next-to-leading terms in
%Eq.~(\ref{gamm2dexp}) lift the degeneracy and single out either of these two states.

In the phase diagrams (Figs.~\ref{pd13} and \ref{pd12}) with interlayer couplings,
we however find different phases. The lower part of the phase diagrams,
%in Fig.~\ref{pd13} and Fig.~\ref{pd12},
where  $J_0$ is as small as $10^{-4}$ and the system looks almost two-dimensional,
shows the fan phase only for $J_3\gtrsim 1.5$ and for $J_2 \gtrsim 5.5$, respectively.
This is because, as described in Sec.~\ref{sec:bseq},
the calculation in two dimensions is intrinsically
based on the assumption that a stable single-magnon condensate
exists as the many-body ground state and the elementary excitations are quasiparticles with
Bogoliubov-like dispersion.
If this assumption is violated, approximating the many-body $T$-matrix with the two-body $T$-matrix calculated
with negative energy cutoff is not justified.
%More in general, the whole dilute magnon gas picture is not guaranteed to be valid.
For example, if a first-order phase transition occurs between the ferromagnetic state (magnon vacuum) and a low-magnetization state (finite density of magnons), evidently the two-body $T$-matrix obtained with the above method will not give any information about the true many-body state below saturation.
Similarly, it will not capture the appearance of two-magnon bound states below  the single-magnon spectrum,
which also breaks the above assumption. Therefore the prediction that either the spiral or the fan phase appear
below the saturation field is reliable only when no instability affects the dilute single-magnon gas picture.
We believe that this is the reason why the phase-separation and bound-state regions
do not appear in the pure 2D analysis.

The other discrepancy between the pure 2D result and the result for
the quasi-2D system is the existence of ``01" phase in the $J_1$-$J_3$ model even for weak $J_0$ (see Fig.~\ref{pd13}).
We note that ``01" phase is forbidden in the $J_0=0$ limit.
A possible reason for this discrepancy is that $J_0$ is not small enough in our calculation.
For small $J_0$, the couplings $\Gamma^{(i)}$ are expanded as $\Gamma^{(i)}= \alpha^{(i)}/|\log |J_0||+
{\cal O}(1/(\log |J_0|)^2)$. Since $1/|\log |J_0||$ decays slowly with decreasing $|J_0|$,
in general the effect of interlayer coupling is still not negligible in this energy scale of $J_0$ near saturation field.
For example, let us consider the boundary between the ``01'' and fan phases in Fig.~\ref{pd13}.
The two relevant couplings for this transition are $\Gamma^{(2)}_1$ and $\Gamma^{(3)}_0$, and the condition
for the phase boundary is given by $\Gamma^{(2)}_1=\Gamma^{(3)}_0$.
Including the next-to-leading order, this equation looks like
\begin{align}
\frac{\alpha^{(2)}_1 } {\left|\log |J_0| \right|} +  \frac{\beta^{(2)}_1 }{\left| \log |J_0| \right|^2}  = \frac{\alpha^{(3)}_0 } {\left|\log |J_0| \right|} +  \frac{\beta^{(3)}_0 }{\left| \log |J_0| \right|^2} .
\label{crossing}
\end{align}
Note that the coefficients of the leading terms, $\alpha^{(2)}_1$ and $\alpha^{(3)}_0$,
are the same as in Eq.~\eqref{alfa1}.
We know  from Fig.~\ref{fig:leading} %Appendix~\ref{bscalc}
that these coefficients are two different regular functions of $J_3'\equiv J_3-1/4$
satisfying $\alpha^{(2)}_1 \ge \alpha^{(3)}_0$.
Assuming that also $\beta^{(2)}_1,\beta^{(3)}_0$ are regular, we can expand Eq.~\eqref{crossing} in powers of $J_3'$ as
\begin{align}
\frac{a}{\left|\log |J_0| \right| } + \left(B+\frac{b}{\left|\log |J_0| \right|} \right) J_3'+\ldots = 0,
\end{align}
%\begin{align}
%\left(A+\frac{a}{\left|\log |J_0| \right| } \right)+ \left(B+\frac{b}{\left|\log |J_0| \right|} \right) J_3'+\ldots = 0
%\end{align}
where $a=\lim_{J_3'\rightarrow0}(\beta^{(2)}_1-\beta^{(3)}_0)$,
$B=\lim_{J_3'\rightarrow0} \partial (\alpha^{(2)}_1-\alpha^{(3)}_0)/\partial J_3'$,
and we have used the relation
$\lim_{J_3'\to 0} (\alpha^{(2)}_1- \alpha^{(3)}_0) =0$.
At small $J_3'$ we can retain only the first two terms in the above equation,
so that there exists an approximate solution for the phase boundary
\begin{align}
J_{3c}' \simeq -\frac{a}{B \left| \log |J_0| \right|}.
\end{align}
Due to this logarithmic singularity the phase boundary can rapidly shift to the $J_3=1/4$ point
with decreasing $|J_0|$.
The ``01''-fan phase boundary in Fig.~\ref{pd13} indeed appears to start this logarithmic behavior for the lower values of $J_0$.
We thus believe that the ``01'' phase will eventually disappear in the pure 2D limit due to this mechanism.
We note that if the two coefficients $\alpha^{(2)}_1$ and $\alpha^{(3)}_0$ were equal for $J_3>1/4$,
we could not obtain this logarithmic singularity, {and in general there is no reason to expect it. We are led to conclude that, given a microscopic model whose ground state energy is in the form of Eq.~\eqref{gse}, the transition between 3D to 2D can be either smooth or singular, according to the value of the leading coefficients $\alpha^{(i)}$ of the effective couplings.}

%when two $\Gamma$s have different coefficients $\alpha$ of $-1/\log |J_0|$ singularity,
%the crossing point $J_{3c}$ of two $\Gamma$s, given by the condition $\Gamma_i=\Gamma_j$, can also have a
%logarithmic singularity as a function of $J_0$.
%The phase boundary between ``01" and fan phases actually appears to scale as
%$J_{3c}-1/4 \sim - a/ \log |J_0|$ in Fig.~\ref{pd13}.

\section{Conclusion and discussion} \label{sec::concl}

In conclusion, we have studied, in a magnon Bose-Einstein condensation picture,
the  triangular $J_1$-$J_2$-$J_3$ antiferromagnet
as a prototypical model
where a combination of competing exchange interactions and geometrical frustration makes
the single-magnon energy minima more-than-doubly degenerate.
We focused on the high applied magnetic field regime, just below the saturation field,
where the two-magnon interaction can be treated quantum-mechanically by means of the dilute Bose gas theory,
and determined the zero-temperature phase diagram as a function of the exchange couplings.
Together with the spiral and fan phases (commonly featured in the phase diagram of helimagnets)
we found an interesting new phase, the ``01'' phase, whose physical properties are in some sense halfway
between the former two.
In the spiral phase, magnons are condensed to a single wave-vector, whereas in both fan and ``01''
phases magnons are condensed to two wave-vectors with an equal density.
While the fan phase is non-chiral, the spiral and ``01'' phases possess chiral order.
The peculiarity of the chiral order in the ``01'' phase is its stripe structure, which results in a novel type of multiferroic. {By studying the singular behavior of the relevant phase boundary as a function of the interlayer coupling $J_0$ for $J_0\to 0$, we explained how the ``01'' phase disappears in the purely 2D limit. This mechanism shows that even a very small interlayer coupling can drastically change the ground state.}

Also, we elucidated the circumstances in which the dilute single-magnon Bose gas picture breaks down;
this occurs quite often when competing ferromagnetic exchange is present\cite{ueda2011nematic}.
%Although our calculation can signal this breakdown, it can not tell about the nature of the true ground state.
%Generally speaking, f
From the discussion in Sec.~\ref{sec:pd_interlayer}, we can expect that the condensed state in such a case
is either a low-magnetization state (with finite density of magnons) or a dilute Bose gas of
multi-magnon bound states.
We can identify the first case by a runaway behavior, i.e., instability, in the dilute Bose gas
theory, which leads to phase separation.
The second case can be captured by the appearance of stable multi-magnon bound states.
In the $J_1$-$J_2$ model, we found that bound two-magnons can have a lower energy than single magnons
around $J_2=1$ with $J_1=-1$
for a weak interlayer coupling regime.
{Exact diagonalization study of the purely two-dimensional model at $J_2 = 1$ with
48 spins indeed indicates a small magnetization jump at
the saturation field and a tendency to the two-magnon
pairing below the magnetization jump. This would correspond to a weak first-order phase transition to spin-nematic state, but this picture needs further confirmation since finite-size effects might be strong in our numerical calculation.}

We also note that the boundary between a dilute Bose gas of magnons (spiral, fan or ``01'') and phase separation
%at fixed $J_0$
in Figs.~\ref{pd13} and \ref{pd12} corresponds to a {\it tricritical point}\cite{chaikin2000principles}
on the $J_3$-$M$  (or  $J_2$-$M$) phase diagram.
%at which a line of second-order phase transition transforms into two first-order phase transition lines.
When phase separation appears, it is accompanied by two first-order phase transitions in the magnetization
($H$-$M$) curve. These two first-order transition lines merge with a line of second-order phase transition at
this phase boundary.
In fact, following the discussion in Sec.~\ref{sec:PS}, if we assume that the sixth-order terms are continuous
and non-vanishing in that neighborhood, the tricritical point is located at $(J_3^*,M_{\rm s})$ [or $(J_2^*,M_{\rm s})$],
where $J_3^*$ ($J_2^*$) denotes the value at which the relevant quartic term
($\Gamma^{(1)}$ for the spiral state, etc.) changes sign and $M_{\rm s}$ the saturated magnetization.

Lastly, let us stress that our analysis  of the ground-state energy of the dilute magnon gas in Sec.~\ref{becdeg}
is quite general  for spin exchange models enjoying a finite degeneracy of single-magnon energy minima.
For example, it is straightforward to introduce  \emph{XXZ}-type spin anisotropy in our model
(see Appendix~\ref{bscalc}).
We performed the analysis for several values of anisotropy, but did not find qualitative modification of
the phase diagrams.
While we have found only single-$q$ or double-$q$ states near saturation in our simple microscopic model,
there is the possibility to find higher-$q$ states with this method by starting from a more complex
Hamiltonian.\footnote{Performing some numerical experiments one can see that states with higher number of modes, up to six, minimize  Eq.~(\ref{gse}) more likely if $\Gamma^{(1)}$ is bigger than other renormalized couplings, namely if there is a sufficient repulsion between magnons at the same wave-vector. However it is difficult to guess when this is the case from a microscopic Hamiltonian.}
We expect, in some complex systems including frustrated interlayer couplings,
that spin anisotropy can drive the transition to interesting multiple-$q$ phases.
%that can be described by an effective pseudo-spin 1/2 Hamiltonian  near some critical magnetic field,
{In fact we note that this is essentially what happens in the spin model for Ba$_3$Mn$_2$O$_8$, which  can accommodate various new phases
including magnetic vortex crystals.\cite{kamiya2013magnetic} }

%** Tricritical point

\begin{acknowledgments}
It is our pleasure to acknowledge stimulating discussions with Ippei Danshita, Akira Furusaki, George Jackeli,
Yoshitomo Kamiya, Yasuyuki Kato,  Tetsuro Nikuni, Tsuyoshi Okubo, Oleg A.\ Starykh,
Shunji Tsuchiya, Hiroaki T.\ Ueda, and especially Daisuke Yamamoto.
GM is supported by a RIKEN FPR fellowship.
This work was supported by KAKENHI No.\ 23540397 from MEXT, Japan.
\end{acknowledgments}

\appendix

\section{Details on the solution of the BS equation} \label{bscalc}

A common method of solving Eq.~(\ref{bseq}) is to make an expansion in lattice harmonics.
Here we briefly describe the method.
First note that, integrating over the Brillouin zone, one has
\begin{align}
& \bra \Gamma(\bq;\bk,\bk') \ket =\non
& 2U \left(1 - \frac{1}{N} \sum_{\bq' \in {\rm BZ}} \frac{\Gamma(\bq';\bk,\bk')}{\epsilon(\bk+\bq')+\epsilon(\bk'-\bq')-2\emin-E} \right),
\label{bsequ}
\end{align}
where $\bra \cdots \ket\equiv (1/N)\sum_{\bq \in BZ}\cdots$ and we have used $\bra \epsilon \ket=0$. For $U\to \infty$ we obtain
\begin{align}
1 - \frac{1}{N} \sum_{\bq' \in {\rm BZ}} \frac{\Gamma(\bq';\bk,\bk')}{\epsilon(\bk+\bq')+\epsilon(\bk'-\bq')-2\emin-E} =0.
\label{bsequinf}
\end{align}
Using Eq.~(\ref{bsequ}), Eq.~(\ref{bseq}) becomes
\begin{align}
& \Gamma(\bq;\bk,\bk') = 2 \epsilon(\bq) +\bra \Gamma(\bk,\bk') \ket \non
& - \frac{1}{N} \sum_{\bq' \in {\rm BZ}} \frac{2 \epsilon(\bq-\bq')}{\epsilon(\bk+\bq')+\epsilon(\bk'-\bq')-2 \emin-E} \Gamma(\bq';\bk,\bk').
\label{bsnou}
	\end{align}

% \subsubsection*{Comments}
At this stage one would like to take a suitable Ansatz for $\Gamma$ and transform the problem,
namely Eqs.~(\ref{bsequinf}) and (\ref{bsnou}), to a linear algebraic system.
The most general expansion in lattice harmonics would contain both harmonics of type sine and cosine.
%be (all coefficients depend on $\bk,\bk'$)
%\begin{align}
%\Gamma(\bq)= \bra \Gamma \ket +\sum_{i=0}^2 \left\{ J_1 A_i \cos {\bf a}_{2i}\cdot \bq + J_3 B_i \cos 2{\bf a}_{2i}\cdot \bq  + J_1 C_i \sin {\bf a}_{2i}\cdot \bq + J_3 D_i \sin 2{\bf a}_{2i}\cdot \bq \right\}
%\label{ansatz}
%\end{align}
%While for $t\geq 0$ (case C-D of Sec.~\ref{sec:spe}) this expansion can actually be simplified, that does not seem the case for our region of interest $-4<t<0$. In fact for $t>0$ the single particle minima are $\bQ_0=(4\pi/3,0), \bQ_3=-\bQ_0$ and taking $\bk,\bk'\in\{\bQ_0,-\bQ_0\}$ one can show that $\{A_i\},\{B_i\},\ldots$ actually do not depend on $i$ (this is a consequence of ${\bf a}_{2i}\cdot \bQ_0=4\pi/3 \; \forall i$). As for $-4<t<0$, the calculation of the $\Gamma$'s in Eqs.~\ref{casebi}-\ref{casebf} does not, in general, enjoy the above simplification.
However, one can get rid of  harmonics of the type sine by considering even functions of $\bq$ to obtain the $\Gamma$'s, namely
\begin{align}
\hat{\Gamma}^{(1)}(\bq) =& \Gamma(\bq;\bQ_0,\bQ_0), \non
 \hat{\Gamma}^{(2)}_1 (\bq) =& \Gamma((\bQ_1-\bQ_0)/2+\bq;\bQ_0,\bQ_1) \non
& +  \Gamma((\bQ_1-\bQ_0)/2-\bq;\bQ_0,\bQ_1), \non
 \hat{\Gamma}^{(2)}_2 (\bq) =& \Gamma((\bQ_2-\bQ_0)/2+\bq;\bQ_0,\bQ_2) \non
&+  \Gamma((\bQ_2-\bQ_0)/2-\bq;\bQ_0,\bQ_2), \non
 \hat{\Gamma}^{(3)} (\bq) = & \Gamma ((\bQ_3-\bQ_0)/2 +\bq;\bQ_0,\bQ_3)  \non
 &+ \Gamma ((\bQ_3-\bQ_0)/2 -\bq;\bQ_0,\bQ_3).
\label{gamma1}
\end{align}
Then the couplings in  the ground-state energy (\ref{gse}) will be given by
\begin{align}
& \Gamma^{(1)} = \hat{\Gamma}^{(1)}(\b0), \non
& \Gamma^{(2)}_1 = \hat{\Gamma}^{(2)}_1 ((\bQ_1-\bQ_0)/2),  \non
& \Gamma^{(2)}_2 = \hat{\Gamma}^{(2)}_2 ((\bQ_2-\bQ_0)/2),  \non
& \Gamma^{(3)}_0 = \hat{\Gamma}^{(3)} ((\bQ_3-\bQ_0)/2),  \non
& \Gamma^{(3)}_1 = \hat{\Gamma}^{(3)} (\bQ_1-(\bQ_3+\bQ_0)/2),  \non
& \Gamma^{(3)}_2 = \hat{\Gamma}^{(3)} (\bQ_2-(\bQ_3+\bQ_0)/2).
\label{hatgamma}
\end{align}
It is possible to verify $ \Gamma^{(3)}_1= \Gamma^{(3)}_2$, as argued previously from symmetry considerations.

In order to solve Eq.~(\ref{gamma1})
we then take the Ansatz Eq.~(\ref{ansatz3}).
%of the type Eq.~(\ref{ansatz}) with $C_i=D_i=0$, namely
%\begin{align}
%\Gamma(\bq)= \bra \Gamma \ket +\sum_{i=0}^2 \left\{ J_1 A_i \cos {\bf a}_{2i}\cdot \bq + J_3 B_i \cos 2{\bf a}_{2i}\cdot \bq  \right\}
%\label{ansatz2}
%\end{align}
For simplicity let us restrict ourselves to $J_2=0$ and define
\begin{widetext}
\begin{align}
& {\bf T}(\bq) = (1,\cos {\bf a}_{0}\cdot \bq, \cos {\bf a}_{2}\cdot \bq, \cos {\bf a}_{4}\cdot \bq, \cos 2{\bf a}_{0}\cdot \bq,\cos 2{\bf a}_{2}\cdot \bq,\cos 2{\bf a}_{4}\cdot \bq, \cos q_z )^T, \\
&\tau_{ij} (\bk,\bk';E) = \frac{1}{N} \sum_{\bq'\in BZ} \frac{T_i(\bq') T_j(\bq')}
{\epsilon\left(\frac{\bk+\bk'}{2}+\bq'\right) +\epsilon\left(\frac{\bk+\bk'}{2}-\bq'\right)
-\epsilon(\bk)-\epsilon(\bk') -E}.
\label{tau}
\end{align}
\end{widetext}
For the 3D calculation in Sec.~\ref{sec:pd_interlayer} we can safely set $E$ to its ``on-shell'' value $E=0$.
However for the 2D calculation in Sec.~\ref{sec:pure2d} ($J_0=0$), a small but finite $E$ will
be used to regularize the integrals in Eq.~(\ref{tau}),
that typically suffer from logarithmic divergences, as explained in Sec.~\ref{sec:bseq}.
%Note that we keep $E$ generic instead of fixing it at  as explained in Section~\ref{sec:bseq}. This will be useful to regularize the integrals in Eq.~(\ref{tau}), that typically suffer from logarithmic divergences in 2D ($J_0=0$).
%This fact is related to the well-known phenomenon of logarithmic vanishing of the effective interaction of the 2D dilute Bose gas \cite{fisher1988dilute}. Physically, an infrared cutoff in momentum space corresponding to the energy scale where the interactions become important (and the bare single particle dispersion becomes inaccurate) is needed. In practice, instead of implementing the momentum cutoff, it is more convenient to adopt the approach of \cite{lee2002energy}, who showed that the cutoff procedure is equivalent (to a good approximation) to taking $E$ to be slightly negative, with amplitude of the order of the chemical potential, namely $E=- C \mu$ ($C$ positive constant).
%More details in Appendix~\ref{2dbose}.

Upon plugging Eq.~(\ref{ansatz3}) into Eq.~(\ref{bsequinf}) and Eq.~(\ref{bsnou}) we derive a system of linear equations for the coefficients in Eq.~(\ref{ansatz3}), that is ${\bf x}=(\vev{\Gamma},A_0, A_1,A_2,C_0,C_1,C_2,D)^T$, which reads
\begin{align}
M {\bf x}= {\bf n} \label{linsys}
\end{align}
with
\begin{widetext}
\begin{align}
& M= \left(
\begin{array}{cccccccc}
\tau_{11} & J_1 \tau_{12} & J_1 \tau_{13} & J_1 \tau_{14} & J_3 \tau_{15} & J_3 \tau_{16} & J_3 \tau_{17} & J_0 \tau_{18}\\
2\tau_{21} & 1+2 J_1 \tau_{22} & 2 J_1 \tau_{23} & 2 J_1 \tau_{24} & 2 J_3 \tau_{25} & 2 J_3 \tau_{26} & 2 J_3 \tau_{27} & 2 {J_0} \tau_{28} \\
2\tau_{31} & 2 J_1 \tau_{32} & 1+ 2 J_1 \tau_{33} & 2 J_1 \tau_{34} & 2 J_3 \tau_{35} & 2 J_3 \tau_{36} & 2 J_3 \tau_{37} & 2 {J_0} \tau_{38}  \\
2\tau_{41} & 2 J_1 \tau_{42} & 2 J_1 \tau_{43} & 1+2 J_1 \tau_{44} & 2 J_3 \tau_{45} & 2 J_3 \tau_{46} & 2 J_3 \tau_{47} & 2 {J_0} \tau_{48} \\
2\tau_{51} & 2 J_1 \tau_{52} & 2 J_1 \tau_{53} & 2 J_1 \tau_{54} & 1+ 2 J_3 \tau_{55} & 2 J_3 \tau_{56} & 2 J_3 \tau_{57} & 2 {J_0} \tau_{58}  \\
2\tau_{61} & 2 J_1 \tau_{62} & 2 J_1 \tau_{63} & 2 J_1 \tau_{64} & 2 J_3 \tau_{65} & 1+2 J_3 \tau_{66} & 2 J_3 \tau_{67} & 2 {J_0} \tau_{68}  \\
2\tau_{71} & 2 J_1 \tau_{72} & 2 J_1 \tau_{73} & 2 J_1 \tau_{74} & 2 J_3 \tau_{75} & 2 J_3 \tau_{76} & 1+2 J_3 \tau_{77} & 2 {J_0} \tau_{78} \\
2\tau_{81} & 2 J_1 \tau_{82} & 2 J_1 \tau_{83} & 2 J_1 \tau_{84} & 2 J_3 \tau_{85} & 2 J_3 \tau_{86} & 2 J_3 \tau_{87} & 1+2 {J_0} \tau_{88} \\
\end{array}
\right), \label{bigmatrix} \\
& {\bf n} =\begin{cases}
(1,2,2,2,2,2,2,2)^T & {\rm for } \; \hat{\Gamma}^{(1)}, \\
\left(2,4 \cos \frac{\bk'-\bk}{2}\cdot {\bf a}_0, 4 \cos \frac{\bk'-\bk}{2}\cdot {\bf a}_2,
4 \cos \frac{\bk'-\bk}{2}\cdot {\bf a}_4, 4 \cos (\bk'-\bk)\cdot {\bf a}_0,  \ldots,
4\cos  (k_z'-k_z)\right)^T & {\rm otherwise}.
\end{cases}
\end{align}
\end{widetext}
The matrix elements can be calculated numerically in three dimensions.

In two dimensions, we  recognize that the integration in   Eq.~\eqref{bsnou} is dominated at small $E$
by the neighborhoods of the solutions of $\epsilon(\bk+\bq')+\epsilon(\bk'-\bq')=0$
(one, two or six solutions depending on the choice of $\bk,\bk'\in \{\bQ_i\}$).
The quantities in Eq.~\eqref{hatgamma} therefore have the singular behavior
\begin{align}
 \hat{\Gamma}^{(1)}(\bq)=\frac{\hat{\alpha}^{(1)}(\bq) } {\left|\log |E| \right|}  + {\cal O} \left(\frac{1}{\left| \log |E| \right|^2} \right)
\end{align}
(and similarly for the others) at small $E$.
By plugging into Eq.~\eqref{bsnou} and keeping only the leading terms
we obtain a set of coupled algebraic equations whose solution is given in the form

\begin{align}
& \alpha^{(1)} \equiv \hat{\alpha}^{(1)}({\bf 0}) =\frac{8\pi}{\sqrt{3}} \sqrt{\det (h^{(1)}/2)}, \non
&\alpha^{(2)}_1 \equiv \hat{\alpha}^{(2)}_1 ((\bQ_1-\bQ_0)/2) =\frac{8\pi}{\sqrt{3}} \sqrt{\det (h^{(2)}_1/2)},  \non
&\alpha^{(2)}_2 \equiv \hat{\alpha}^{(2)}_2 ((\bQ_2-\bQ_0)/2) =\frac{8\pi}{\sqrt{3}} \sqrt{\det (h^{(2)}_2/2)},  \non
&\alpha^{(3)}_0\equiv \hat{\alpha}^{(3)}(\bQ_3) = \frac{8\pi}{\sqrt{3}} \sqrt{\det (h^{(3)}/2)},  \non
& \alpha^{(3)}_1\equiv \hat{\alpha}^{(3)}(\bQ_1) =0, \non
& \alpha^{(3)}_2\equiv \hat{\alpha}^{(3)}(\bQ_2) =0.
\end{align}
Here we have introduced the relevant Hessian matrices
\begin{widetext}
\begin{align}
& h^{(1)}_{ij} =\partial_{k_i} \partial_{k_j} [\epsilon(\bQ_0+\bk)+\epsilon(\bQ_0-\bk)]  _{\bk={\bf 0}}, \non
& h^{(2)}_{1,ij} =\partial_{k_i} \partial_{k_j} \left[\epsilon \left(\frac{\bQ_0+\bQ_1}{2}+\bk \right)+\epsilon \left(\frac{\bQ_0+\bQ_1}{2}-\bk \right) \right]  _{\bk=\frac{\bQ_1-\bQ_0}{2}}, \non
& h^{(2)}_{2,ij} =\partial_{k_i} \partial_{k_j} \left[\epsilon \left(\frac{\bQ_0+\bQ_2}{2}+\bk \right)+\epsilon \left(\frac{\bQ_0+\bQ_2}{2}-\bk \right) \right]  _{\bk=\frac{\bQ_2-\bQ_0}{2}}, \non
& h^{(3)}_{ij} =\partial_{k_i} \partial_{k_j} \left[2\epsilon \left(\bk \right) \right]  _{\bk=\bQ_0} = h^{(1)}_{ij}.
\end{align}
\end{widetext}
It is easy to verify that $\det h^{(1)}=\det h^{(3)}$ and $\det h^{(2)}_1=\det h^{(2)}_2$.
Equations~\eqref{alfa1} and \eqref{alfa2} follow.

%\textcolor{red}{In two dimensions we can extract the leading term for small $E$ in the following way.
%We choose $\bk,\bk'\in \{\bQ_i\}$ and note that the integrals in Eq.~(\ref{tau}) are dominated by the contribution of the neighborhood(s)  $\bq'\simeq  \pm (\bk'-\bk)/2$ (and also $\pm R_{\pi/3} (\bk'-\bk)/2,\pm R_{2\pi/3} (\bk'-\bk)/2$ if $\bk=-\bk'$), where the denominator is approximately $(1/2) h_{mn} p_m p_n-E$, with $\bp=\bq'- (\bk'-\bk)/2 $ (or $\bp=\bq'- R_{\pi/3}(\bk'-\bk)/2 $, etc.) and $h_{mn}$ the Hessian matrix at $\bp=0$.
%Thus $\tau_{ij}=a_{ij}-b_{ij} \log |E| +{\cal O}(E)$.
%The coefficients $b_{ij}$ of the leading term can be calculated analytically.
%Plugging this form into Eq.~(\ref{linsys})
%one can obtain the leading term of the $\Gamma$'s, which are of order $-1/\log |E|$,
%as reported in Sec.~\ref{sec:pure2d}. %Beyond the leading term a numerical computation is needed.
%}

Note that in order to include a generic \emph{XXZ}-type spin anisotropy,
namely to change the various terms in the Hamiltonian Eq.~(\ref{micr}) from
$J_\alpha \mathbf{S}_i\cdot\mathbf{S}_j$ to $J_\alpha (S^x_iS^x_j+S^y_iS^y_j+\Delta_\alpha S^z_iS^z_j)$,
one just need the replacement $J_\alpha \to J_\alpha\Delta_\alpha$ in Eqs.~(\ref{ansatz3}) and (\ref{bigmatrix}).

%\bibliography{magnon}
\bibliography{triangJ1J3_arxiv4.bbl}

\end{document}